\documentclass[11pt]{article}
\usepackage{}
\usepackage{amssymb}
\usepackage{amsfonts}
\usepackage{mathrsfs}
\usepackage{graphicx}
\usepackage{amsmath}
\usepackage{color}
\usepackage{amsthm}
\usepackage{amsmath}
\usepackage{pifont,bm}
\usepackage{tikz}
\usepackage{enumerate}

\theoremstyle{definition}

\makeatletter
\def\@biblabel#1{[#1]}
\makeatother

\makeatletter \@addtoreset{equation}{section}

\begin{document}

\begin{titlepage}
\title{\bf{Characteristics of rogue waves in the scalar and vector nonlocal nonlinear Schr\"{o}dinger equations
\footnote{This work is supported by NSFC under Grant Nos. 12201622 and 11975306.\protect\\
\hspace*{3ex}$^{*}$Corresponding authors.\protect\\
\hspace*{3ex} E-mail addresses:xiubinwang@163.com, xbwang@cumt.edu.cn, sftian@cumt.edu.cn}
}}
\author{Xiu-Bin Wang, Shou-Fu Tian$^{*}$\\
\small \emph{School of Mathematics, China University of Mining and Technology, Xuzhou 221116,} \\
\small \emph{People's Republic of China}\\
\date{}}
\thispagestyle{empty}
\end{titlepage}
\maketitle

\vspace{-0.5cm}
\begin{center}
\rule{15cm}{1pt}\vspace{0.3cm}

\parbox{15cm}{\small
{\bf Abstract}\\
\hspace{0.5cm} In this work,
general higher-order rogue wave solutions of the parity-time symmetric scalar and vector nonlocal nonlinear Schr\"{o}dinger equations are calculated theoretically via a DDT by a separation of variable technique.
Furthermore, in order to understand these solutions better, the main characteristics of the obtained solutions are discussed clearly and conveniently.
Our results show that the dynamics of these solutions exhibits rich
patterns, most of which have no counterparts in the corresponding local equations.
}

\vspace{0.5cm}
\parbox{15cm}{\small{

\vspace{0.3cm} \emph{Key words:} The scalar and vector nonlocal nonlinear Schr\"{o}dinger equations; Rogue waves; Darboux-dressing transformation (DDT);
A variable separation technique.

 } }
\end{center}
\vspace{0.3cm} \rule{15cm}{1pt} \vspace{0.2cm}

\section{Introduction}

It is well-known that integrable nonlinear systems play an important role in the field of mathematical physics.
Most of these integrable nonlinear systems are local equations. In other words, the solutions' evolution
relies only on the local solution value.
In recent years, integrable nonlocal nonlinear systems have attracted a lot of attention and have been
studied extensively.
This type of equation is parity-time symmetric because it is invariant under the joint transformation and complex conjugation.
The first such system was introduced by Ablowitz and Musslimani in 2013 \cite{MJAAZH-2013,MJAAZH-2018,MJAAZH-2014}
\begin{equation*}\label{ABNLS}
\textrm{i}q_{t}(x,t)+q_{xx}(x,t)\pm 2q^2(x,t)\bar{q}(-x,t)=0,
\end{equation*}
where $\pm$ determines whether the above equation is focused or defocused and the overline denotes the complex conjugation.
It is worth mentioning that parity-time symmetric equations play a vital role in optics and other physical fields recently \cite{VVJJ-2016}.
Following the above nonlocal parity-time symmetric nonlinear equation,
some new reverse space-time and reverse-time type nonlocal nonlinear integrable equations were also quickly proposed and studied over the past few years \cite{MJJZHM2017,1MJJZHM2018,2MJJZHM2018,LSYLS1,WXBTS2,MWX11,MWX133,MWX22}.

Rogue waves originally attracted a lot of attention due to the mysterious and severely destructive oceanic surface waves \cite{RWWWPE2008,RWKCPE2008}.
This types of waves are spontaneous large waves that ``appear from nowhere and disappear with no trace'' \cite{AKNA-2009}.
The first analytical expression of rogue wave was derived for the standard nonlinear Schr\"{o}dinger equation (NLSE) by Peregrine \cite{DHPPPP1983}.
After that, higher-order rogue waves in the
NLSE were found, and their interesting dynamical patterns were discussed \cite{AKHNA-2009,AAHNA-2010,KEDJA-2011,GBLLL-2012,OYYJK-2012,DPMVB-2013,BFCMD-2013,XSWHJ-2011,GBLLQ-2013,QZYSIAM-2015}.
Nowadays, rogue waves have been rapidly overspread to many research fields
encompassing oceanography \cite{KDHE-2008}, nonlinear optics \cite{DRCR-2007}, Bose-Einstein condensation \cite{YVBV-2009},
superfluid helium \cite{LCZZ-2013}, plasmas \cite{RWFF-2013} and even finance \cite{YZYS-2010}.
As an unexplored and interesting subject, rogue waves in nonlocal integrable systems have received much attention recently \cite{YJKLS-2019,YJKLS-2020,WXB-2022,HEJSLS2020,HEJSLS2027}.

In this work, we focus on a scalar nonlocal reverse-time NLSE
\begin{equation}\label{RTNLSE}
\textrm{i}\psi_{t}(x,t)+\psi_{xx}(x,t)+2\psi^2(x,t)\psi(x,-t)=0,
\end{equation}
and a vector nonlocal reverse-time NLSE
\begin{align}\label{RTNLSE2}
&\textrm{i}\widetilde{\psi}_{t}(x,t)+\widetilde{\psi}_{xx}(x,t)+2\widetilde{\psi}(x,t)\widetilde{\psi}^{\mbox{T}}(x,-t)\widetilde{\psi}(x,t)=0,\notag\\
&\widetilde{\psi}=(\psi_{1},\psi_{2})^{\mbox{T}}.
\end{align}
Eq.\eqref{RTNLSE} and Eq.\eqref{RTNLSE2} are completely integrable and have been investigated extensively \cite{MJJZHHH-2017,NLSRTT-2020,JKYPLA-2019,YERWSAMP-2020,AMLMAW-2019,LCZLS-2020,CYLYB-2021}.
For example, Yang derived both the bounded and collapsing soliton solutions for Eq.\eqref{RTNLSE}
by using the inverse scattering transform \cite{JKYPLA-2019},
Zhang et al. presented the general soliton solutions for Eq.\eqref{RTNLSE} via the binary  Darboux transformation (DT) \cite{YERWSAMP-2020}.
More recently, Ma generalized Eq.\eqref{RTNLSE} into the multicomponent case and derived the $N$-soliton solution of Eq.\eqref{RTNLSE} with vanishing
background  via the Riemann-Hilbert approach \cite{AMLMAW-2019}.
Thanks to the generalized
DT, rogue wave solutions for Eq.\eqref{RTNLSE} and Eq.\eqref{RTNLSE2} have been obtained recently \cite{LCZLS-2020,CYLYB-2021}.
Inspired by previous works, an improved version of DT is derived in this work.
This new separation of variable technique will allow the rogue waves to be obtained
without any calculations of the derivatives.
This technique will provide us facilities for
construction of rogue waves using generalized DT.

The purpose of this work is to develop a variable separation technique and to solve a family of solutions of Lax system of Eq.\eqref{RTNLSE} and Eq.\eqref{RTNLSE2}.
Then we trigger the strategy to construct their $N$th-order explicit rogue wave solutions.
Furthermore, the dynamics of these rogue wave solutions obtained in this paper are discussed clearly and conveniently by different choices of free parameters.
More importantly,
the solution dynamics exhibits rich patterns, most of which are different from the previous results.

This paper is organized as follows.
In section 2, we present
the Lax pair and asymptotic expansion of the DDT for Eq.\eqref{RTNLSE}.
Then we employ the variable separation technique to construct
$N$th-order rogue wave solutions solutions of Eq.\eqref{RTNLSE} using a DDT.
Moreover, we present from first to third order rogue wave solutions and illustrate their dynamic behaviors.
In section 3, the variable separation technique in presented in section 2 is used to derive the $N$th-order rogue wave solution for Eq.\eqref{RTNLSE2},
and dynamic behaviors of two lowest rogue wave solutions is displayed graphically.
Finally, the summary of these results is provided in section 4.

\section{A scalar nonlocal nonlinear Schr\"{o}dinger equation}
In this section, the $N$th-order rogue wave solutions of Eq.\eqref{RTNLSE} will be calculated theoretically via a DDT associated with a novel expansion technique.

\subsection{Asymptotic expansion of Darboux-dressing transformation}

Eq.\eqref{RTNLSE} admits the following Lax pair
\begin{equation}\label{DDT-1}
\Psi_{x}=\textbf{U}\Psi,~~
\Psi_{t}=\textbf{V}\Psi,
\end{equation}
where
\begin{equation*}\label{DDT-2}
\left\{ \begin{aligned}
&\textbf{U}=\textrm{i}\lambda\sigma_{3}+Q,\\
&\textbf{V}=2\textrm{i}\lambda^2\sigma_{3}+2\lambda Q+\textrm{i}\sigma_{3}\left(Q^2-Q_{x}\right),
                     \end{aligned} \right.
\end{equation*}
and
\begin{equation*}\label{DDT-3}
Q=\left[
    \begin{array}{cc}
      0 & -\psi(x,-t) \\
      \psi(x,t) & 0  \\
    \end{array}
  \right],~~\sigma_{3}=\left[
                     \begin{array}{cc}
                       1 & 0 \\
                       0 & -1 \\
                     \end{array}
                   \right],
\end{equation*}
with the spectral parameter $\lambda$.
By using the compatibility condition of system \eqref{DDT-1}
\begin{equation*}\label{CCSS1}
\textbf{U}_{t}-\textbf{V}_{x}+[\textbf{U},\textbf{V}]=0,
\end{equation*}
one can derive directly Eq.\eqref{RTNLSE}, where commutator $[\mathbb{A},\mathbb{B}]=\mathbb{A}\mathbb{B}-\mathbb{B}\mathbb{A}$.
The Lax pair \eqref{DDT-1} admits the following symmetric condition
\begin{equation*}\label{DDT-4}
\textbf{U}(x,-t;-\lambda)=-\textbf{U}^{\mbox{T}}(x,t;\lambda),~~\textbf{V}(x,-t;-\lambda)=\textbf{V}^{\mbox{T}}(x,t;\lambda).
\end{equation*}
We know that for $\lambda\in\mathbb{C}$, $\Phi_{1}=\Psi_{1}^{\mbox{T}}(x,-t)$ solves the adjoint eigenvalue problem
\begin{equation*}\label{DDT-5}
\Phi_{x}=-\Phi \textbf{U}(Q,\lambda),~~\Phi_{t}=-\Phi \textbf{V}(Q,\lambda),
\end{equation*}
at $\lambda=-\lambda_{1}$, while $\Psi_{1}$ is a solution of the linear matrix eigenvalue problem \eqref{DDT-1} at $\lambda=\lambda_{1}$.

A suitable DDT for Eq.\eqref{RTNLSE} is given by
\begin{equation}\label{DDT-6}
\Psi[1]=\textbf{D}\Psi,~~\textbf{D}=\textbf{I}_{2\times2}-\frac{2\lambda_{1}}{\lambda+\lambda_{1}}\textbf{P},~~
\textbf{P}=\frac{\Delta[0](x,t)\Delta[0]^{\mbox{T}}(x,-t)}{\Delta[0]^{\mbox{T}}(x,-t)\Delta[0](x,t)},
\end{equation}
in which $\textbf{I}_{2\times2}=\mbox{diag}(1,1)$, $\Delta[0]=[\varphi_{0},\varphi_{1}]^{\mbox{T}}$, and
$\Psi(x,t;\lambda_{1})$ is a fundamental solution for Eq.\eqref{DDT-1} corresponding to $\lambda=\lambda_{1}$.
Next, it is useful to note that the DDT \eqref{DDT-6} can
be replaced with the alternative form
\begin{equation}\label{DDT-8}
\Psi[1]=\textbf{T}\Psi,~~\textbf{T}=(\lambda+\lambda_{1})\textbf{I}_{2\times2}-2\lambda_{1}\textbf{P}.
\end{equation}
Because $\textbf{T}$ is also a DT of Eq.\eqref{RTNLSE}, it follows that
\begin{equation}\label{DDT-9}
\textbf{T}_{x}+\textbf{T}\textbf{U}+\textbf{U}_{1}\textbf{T}=0,~~\textbf{T}_{t}+\textbf{T}\textbf{U}+\textbf{V}_{1}\textbf{T}=0.
\end{equation}
The matrices $\textbf{U}_{1}$ and $\textbf{V}_{1}$ are obtained by replacing $Q$ with $Q_{1}$ in $\textbf{U}$ and $\textbf{V}$, respectively.
It follows from \eqref{DDT-8} and \eqref{DDT-9}  that
\begin{equation*}\label{DDT-10}
Q_{1}=Q_{0}+2\textrm{i}\lambda_{1}[\sigma_{3},\textbf{P}].
\end{equation*}
Here, we note that
\begin{equation*}\label{DDT-11}
\textbf{T}|_{\lambda=\lambda_{1}}\Delta[0]=0.
\end{equation*}
It means that the DDT \eqref{DDT-8} cannot be iterated continuously for the same spectral parameter. In order to eliminate this limitation, we introduce the following expansion theorem which can be used to produce new
solutions for the same spectral parameter.
\\

\noindent
\textbf{Theorem 2.1}
Let $\Psi(\lambda)|_{\lambda=\lambda_{1}(1+\epsilon)}$  be a solution of the Lax system \eqref{DDT-1} corresponding to the spectral parameter
$\lambda_{1}(1+\epsilon)$ and a seed solution $\psi^{[0]}$.
If $\Psi(\lambda)$ has an expansion at $\lambda_{1}$
\begin{equation*}\label{DDT-12}
\Psi(\lambda)|_{\lambda=\lambda_{1}(1+\epsilon)}=\Psi_{0}\epsilon+\Psi_{0}\epsilon+\Psi_{0}\epsilon^2+\cdots,
\end{equation*}
where
\begin{align*}\label{DDT-13}
&\Delta[n]=\left[
            \begin{array}{c}
              \varphi_{0}^{[n]} \\
              \varphi_{1}^{[n]} \\
            \end{array}
          \right]=\lambda_{1}\Delta[n-1]+\textbf{T}[n]\Upsilon[n-1],~~n\geq1,\notag\\
&\Delta[0]=\left[
            \begin{array}{c}
              \varphi_{0}^{[0]} \\
              \varphi_{1}^{[0]} \\
            \end{array}
          \right]=\Psi_{0},\notag\\
&\Upsilon[n-1]=\Delta[n-1](\Psi_{j}\rightarrow\Psi_{j+1}),~~j=0,1,2,\cdots,
\end{align*}
and
\begin{equation*}\label{DDT-14}
\textbf{T}[n]=2\lambda_{1}(\textbf{I}_{2\times2}-\textbf{P}[n]),~~
\textbf{P}[n]=\frac{\Delta[n-1](x,t)\Delta^{\mbox{T}}[n-1](-x,t)}{\Delta^{\mbox{T}}[n-1](-x,t)\Delta[n-1](x,t)},
\end{equation*}
are solutions of the Lax system \eqref{DDT-1} corresponding to
$\lambda_{1}$ and solution $\psi^{[n]}$
\begin{align*}\label{DDT-15}
\psi^{[n]}=\psi^{[n-1]}+\frac{4\textrm{i}\lambda_{1}\varphi_{0}^{[n-1]}(x,-t)\varphi_{1}^{[n-1]}(x,t)}
{\varphi_{0}^{[n-1]}(x,t)\varphi_{0}^{[n-1]}(x,-t)+\varphi_{1}^{n-1}(x,t)\varphi_{1}^{[n-1]}(x,-t)}.
\end{align*}

\noindent
\textbf{Proof}: Similar to \cite{QZYSIAM-2015}, the proof of this theorem can be given by using mathematical induction.

\subsection{The variable separation technique}

Just as in the case of local NLSE \cite{GBLLL-2012}, we start with the plane wave solution of Eq.\eqref{RTNLSE}
\begin{equation}\label{DDT-71}
\psi^{[0]}=\rho\exp\left(2\textrm{i}\rho^2 t\right),
\end{equation}
where $\rho$ is free constant.
Then we find a family of the solutions of the Lax system \eqref{DDT-71} corresponding to the spectral parameter $\lambda$ in the following form
\begin{equation}\label{DDT-81}
\Psi=\left[
       \begin{array}{c}
         \varphi_{0} \\
         \varphi_{1} \\
       \end{array}
     \right]=
\Lambda\mathcal {R}\mathcal {E}\mathcal {Z},~~\mathcal {R}=\exp(\textrm{i}\Theta x),~~\mathcal {E}=\exp(\textrm{i}\Omega t),
\end{equation}
where
\begin{equation*}\label{DDT-91}
\Lambda=\left[
          \begin{array}{cc}
            \exp\left(-\textrm{i}\rho^2 t\right) & 0 \\
            0 & \exp\left(\textrm{i}\rho^2 t\right)\\
          \end{array}
        \right],
\end{equation*}
with an arbitrary complex vector $\mathcal {Z}$.
Here, the two matrices $\Theta$ and $\Omega$ must satisfy
\begin{equation}\label{DDT-110}
[\Theta,\Omega]=\Theta\Omega-\Omega\Theta=0.
\end{equation}
Putting \eqref{DDT-81} into \eqref{DDT-1} reaches to
\begin{equation}\label{DDT-111}
\Lambda_{x}+\textrm{i}\Lambda\Theta-\textbf{U}\Lambda=0,~~\Lambda_{t}+\textrm{i}\Lambda\Theta-\textbf{V}\Lambda=0.
\end{equation}
Solving the above conditions \eqref{DDT-110} and \eqref{DDT-111}, we obtain
\begin{equation*}\label{DDT-12}
\Theta=\left[
         \begin{array}{cc}
           \lambda & \textrm{i}\rho \\
           -\textrm{i}\rho & -\lambda \\
         \end{array}
       \right],~~\Omega=\Theta^2+2\lambda\Theta-\left(\lambda^2+\rho^2\right).
\end{equation*}
Then the exponential matrices $\mathcal {R}$ and $\mathcal {E}$ in \eqref{DDT-81} can be written as
\begin{equation}\label{DDT-131}
\mathcal {R}=\frac{1}{\tau}\left[
                             \begin{array}{cc}
                               \Theta_{1} & \Theta_{2}  \\
                               \Theta_{3} & \Theta_{4}  \\
                             \end{array}
                           \right],~~
\mathcal {E}=\frac{1}{\xi}\left[
                             \begin{array}{cccc}
                               \Omega_{1} & \Omega_{2} \\
                               \Omega_{3} & \Omega_{4} \\
                             \end{array}
                           \right],
\end{equation}
where
\begin{align*}\label{DDT-141}
&\Theta_{1}=\tau\cos(\tau x)+\textrm{i}\lambda\sin(\tau x),~~
\Theta_{3}=-\Theta_{2}=\rho\sin(\tau x),\notag\\
&\Theta_{4}=\tau\cos(\tau x)-\textrm{i}\lambda\sin(\tau x),\notag\\
&\Omega_{1}=\xi\cos(\xi t)+2\textrm{i}\lambda^2\sin(\xi t),~~
\Omega_{3}=-\Omega_{2}=2\lambda\rho \sin(\xi t),\notag\\
&\Omega_{4}=\xi\cos(\xi t)-2\textrm{i}\lambda^2\sin(\xi t),~~\xi=2\lambda\tau,\notag\\
&\tau=\sqrt{\lambda^2+\rho^2}.
\end{align*}

\subsection{Construction of $N$th-order rogue wave solutions}

In what follows, we construct $N$th-order rogue waves of Eq.\eqref{RTNLSE} using the Theorem 2.1 presented in the previous subsection.
Taking $\lambda=\textrm{i}\rho(1+\epsilon)$ in \eqref{DDT-131}.
Then using Taylor series expansions for the trigonometric and exponential functions,
the matrix $\mathcal {R}$ has the expansion at $\epsilon=0$ as
\begin{equation*}\label{DDT-15}
\mathcal {R}\big|_{\lambda=\textrm{i}\rho(1+\epsilon)}=\sum_{n=1}^{\infty}\mathcal {R}_{n}\epsilon^n,
\end{equation*}
where
\begin{equation*}\label{DDT-16}
\mathcal {R}_{n}=\left[
                            \begin{array}{cc}
                              \alpha_{n}-\beta_{n}-\beta_{n-1} & -\beta_{n}  \\
                              \beta_{n} & \alpha_{n}+\beta_{n}+\beta_{n-1} \\
                            \end{array}
                          \right],
\end{equation*}
with
\begin{equation*}\label{DDT-17}
\left\{ \begin{aligned}
&\alpha_{n}=\sum_{l=0}^{\left[\frac{n}{2}\right]}\textbf{C}_{n-l}^{l}2^{n-2l}\textbf{A}_{2(n-l)},\\
&\beta_{n}=\sum_{l=0}^{\left[\frac{n}{2}\right]}\textbf{C}_{n-l}^{l}2^{n-2l}\textbf{A}_{2(n-l)+1},\\
&\textbf{C}_{n}^{m}=\frac{n!}{m!(n-m)!},
~~\textbf{A}_{m}=\frac{\rho^mx^{m}}{m!},~~n\geq m, ~~n,m\in\mathbb{N}^{+}.
                     \end{aligned} \right.
\end{equation*}
Following the same way, the matrix $\mathcal {G}$ has the expansion at $\epsilon=0$ as
\begin{equation*}\label{DDT-18}
\mathcal {E}\big|_{\lambda=\textrm{i}\rho(1+\epsilon)}=
\sum_{n=0}^{\infty}\mathcal {E}_{n}\epsilon^{n},
\end{equation*}
where
\begin{equation*}\label{DDT-19}
\mathcal {E}_{n}=\left[
                            \begin{array}{cccc}
                              \gamma_{n}-\textrm{i}\theta_{n}-\textrm{i}\theta_{n-1} & -\gamma_{n}  \\
                              \gamma_{n} & \gamma_{n}+\textrm{i}\theta_{n}+\textrm{i}\theta_{n-1}  \\
                            \end{array}
                          \right],
\end{equation*}
with
\begin{equation*}\label{DDT-20}
\left\{ \begin{aligned}
&\gamma_{n}=\sum_{l=0}^{\left[\frac{3n}{4}\right]}\sum_{m=0}^{l}(-1)^{n-l}\textbf{C}_{n-l}^{m}
\textbf{C}_{2(n-l)}^{l-m}2^{n-l-m}\textbf{B}_{2(n-l)},\\
&\theta_{n}=\sum_{l=0}^{\left[\frac{3n+1}{4}\right]}\sum_{m=0}^{l}(-1)^{n-l}\textbf{C}_{n-l}^{m}
\textbf{C}_{2(n-l)+1}^{l-m}2^{n-l-m}\textbf{B}_{2(n-l)+1},\\
&\textbf{B}_{m}=\frac{\rho^{2m}2^{m}t^{m}}{m!},~~l\in\mathbb{N}^{+}.
                     \end{aligned} \right.
\end{equation*}
Let us next assume $\omega_{k}$ to be an arbitrary polynomial function of $\epsilon$ given by
\begin{equation*}\label{DDT-21}
\mathcal {Z}_{0}(\epsilon)=\sum_{k=0}^{n}\omega_{k}\epsilon^k,~~
\omega_{k}=\left[
                      \begin{array}{c}
                        \omega_{1,k} \\
                        \omega_{2,k}\\
                      \end{array}
                    \right],
\end{equation*}
thus solution \eqref{DDT-8} has an expansion
\begin{align*}\label{DDT-22}
\Psi\big|_{\lambda=\textrm{i}\rho(1+\epsilon)}=\sum_{n=0}^{\infty}\Psi_{n}\epsilon^{n},~~
\Psi_{n}=\Lambda\sum_{k=0}^{n}
\sum_{j=0}^{n}\mathcal {F}_{k}\mathcal {G}_{j}\omega_{n-k-j}.
\end{align*}
Then taking $\lambda_{1}=\textrm{i}\rho$ in Theorem 2.1 reaches to the $N$th-order rogue  wave solutions of Eq.\eqref{RTNLSE}.
Here it is necessary to emphasize that the $N$th-order rogue  waves solutions presented in this work must be not an even function of $t$ and does not satisfy the corresponding local NLSE.
%
%

According to the above results, we next analyze the dynamic behaviors of the rogue wave solutions in the framework of Eq.\eqref{RTNLSE} by graphic representations.

(I) Taking $N=1$, we have the first-order rogue wave solution of Eq.\eqref{RTNLSE}. Figure 1 is plotted for the first-order rogue wave $|\psi|$ for Eq.\eqref{RTNLSE} with suitable parameters, which is localized both in time and space, thus revealing the usual rogue wave features.

(II) Taking $N=2$, we have the second-order rogue wave solutions of Eq.\eqref{RTNLSE}. More interesting are the collapsing solutions,
as observed in Figure 2, we find that the wave contains six singular peaks, which are arranged in circular pattern.

(III) In order to exhibit the effectiveness of our results, we discuss the third-order rogue wave solutions graphically.
Taking $N=3$, we have the third-order rogue wave solutions of Eq.\eqref{RTNLSE}.
Figure 3 containing ten
singular peaks surrounding one Peregrine-like nonsingular peak.
Figure 4 containing twelve singular peaks, which are arranged in two circular patterns.
To the best of our knowledge, the similar phenomena have been not reported in the local NLSEs.

$~~~~~~~~~~~~~~~~~~~$
{\rotatebox{0}{\includegraphics[width=8.2cm,height=4.8cm,angle=0]{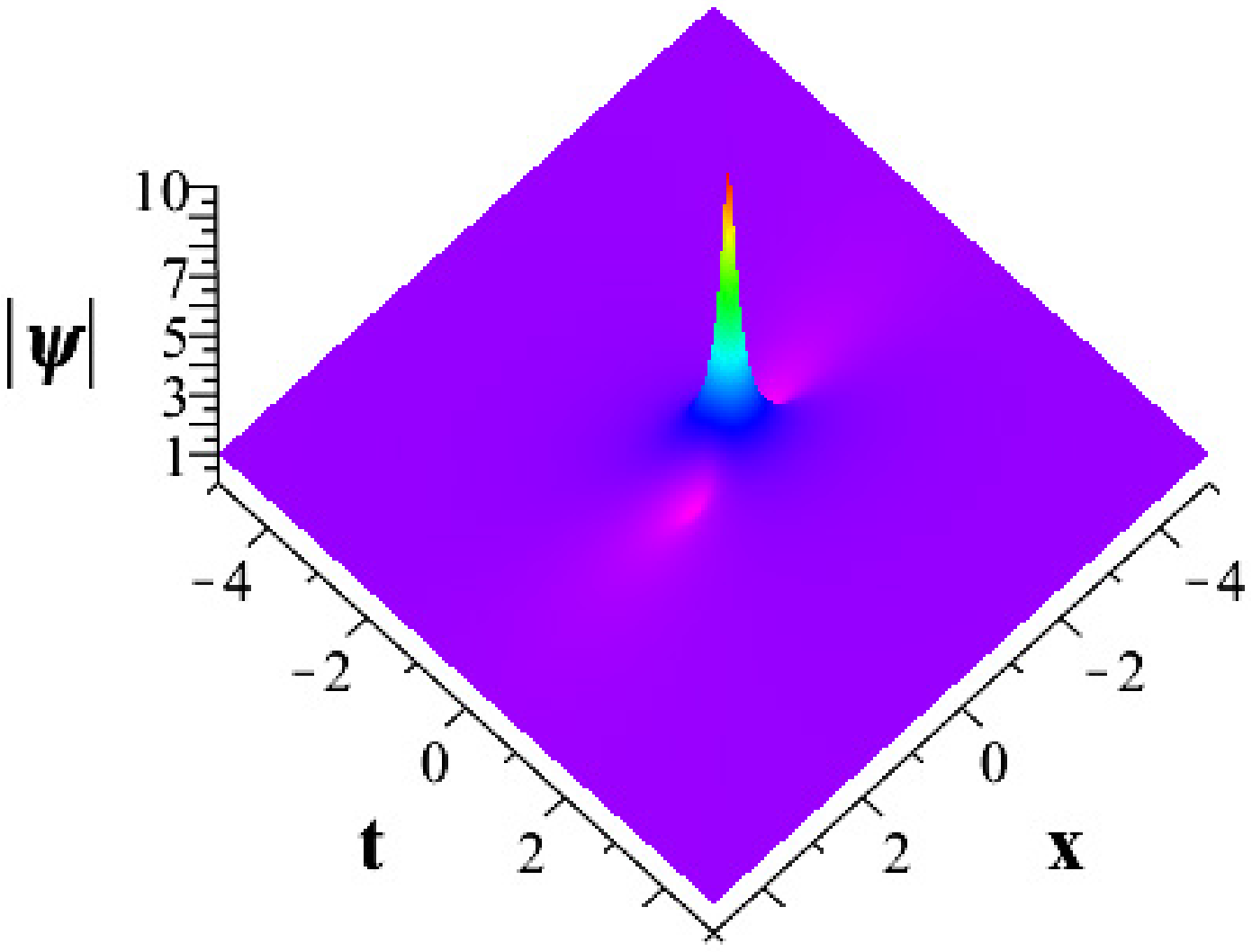}}}\\
$~~~~~~~~~~~~~~~~~~~~~~~~~~~~~~~~~~~~~~~~~~~~~~~~~~~~~~(\textbf{a})$\\

$~~~~~~~~~~~~~~~~~~~~~~~$
{\rotatebox{0}{\includegraphics[width=6.2cm,height=5.6cm,angle=0]{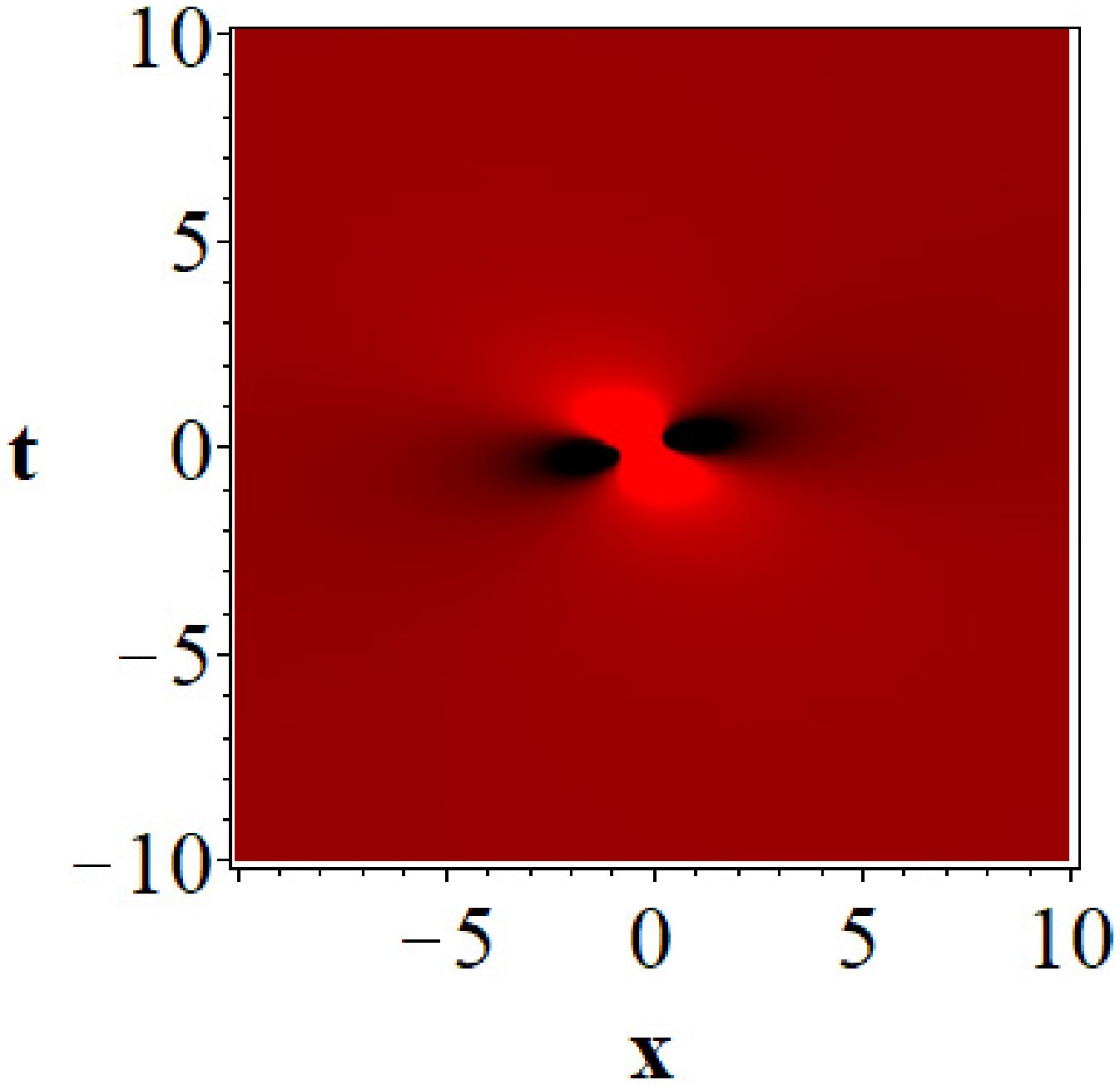}}}\\
$~~~~~~~~~~~~~~~~~~~~~~~~~~~~~~~~~~~~~~~~~~~~~~~~~~~~~~(\textbf{b})$\\

\noindent { \small \textbf{Figure 1.} First-order rogue wave of a scalar nonlocal NLSE
with parameter values
$\rho=1, (\omega_{1,0},\omega_{2,0})=(1,2\textrm{i})$.\\}

$~~~~~~~~~~~~~~~~~~~$
{\rotatebox{0}{\includegraphics[width=8.2cm,height=4.8cm,angle=0]{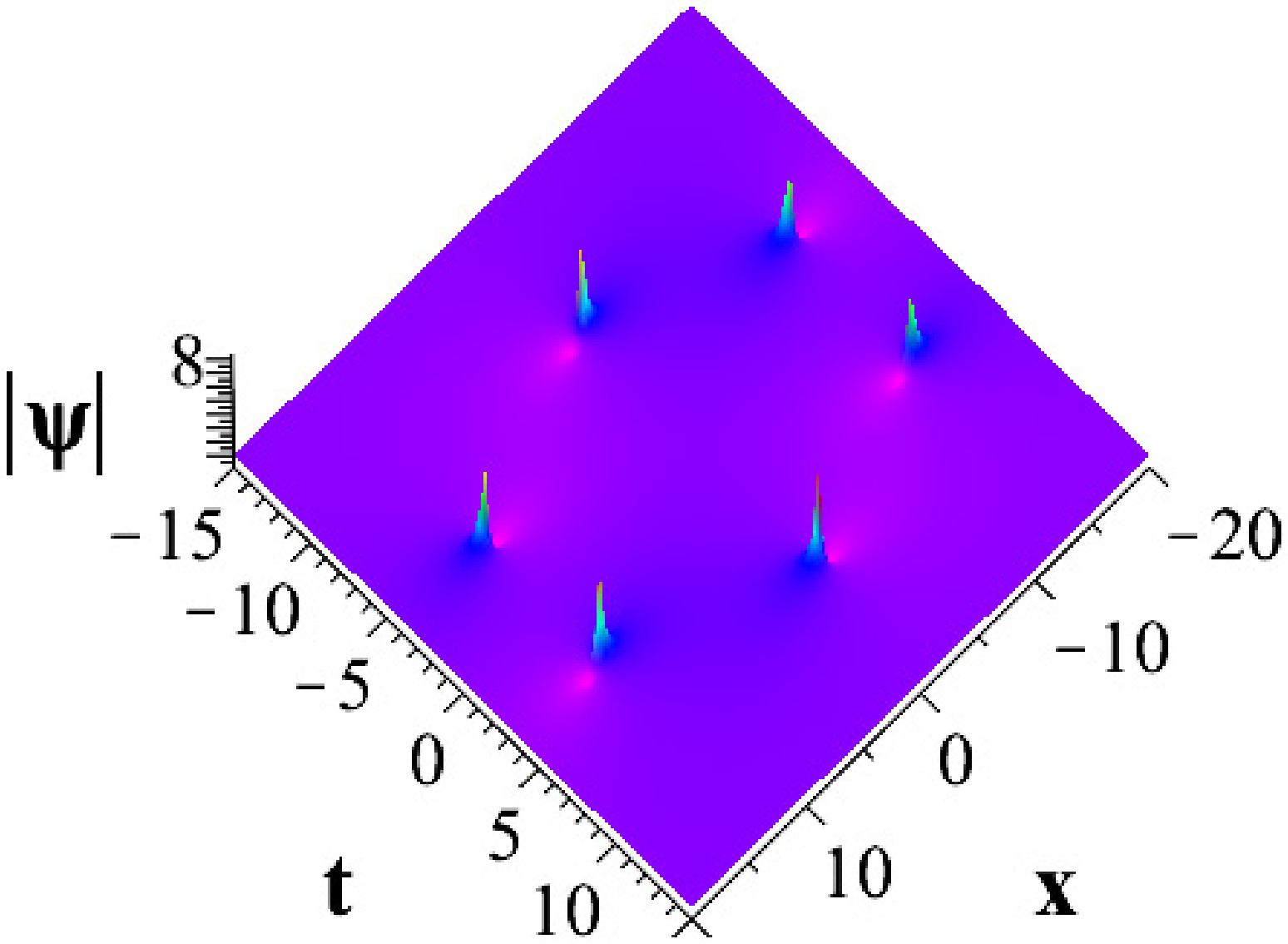}}}\\
$~~~~~~~~~~~~~~~~~~~~~~~~~~~~~~~~~~~~~~~~~~~~~~~~~~~~~~(\textbf{a})$\\

$~~~~~~~~~~~~~~~~~~~~~~~$
{\rotatebox{0}{\includegraphics[width=6.2cm,height=5.6cm,angle=0]{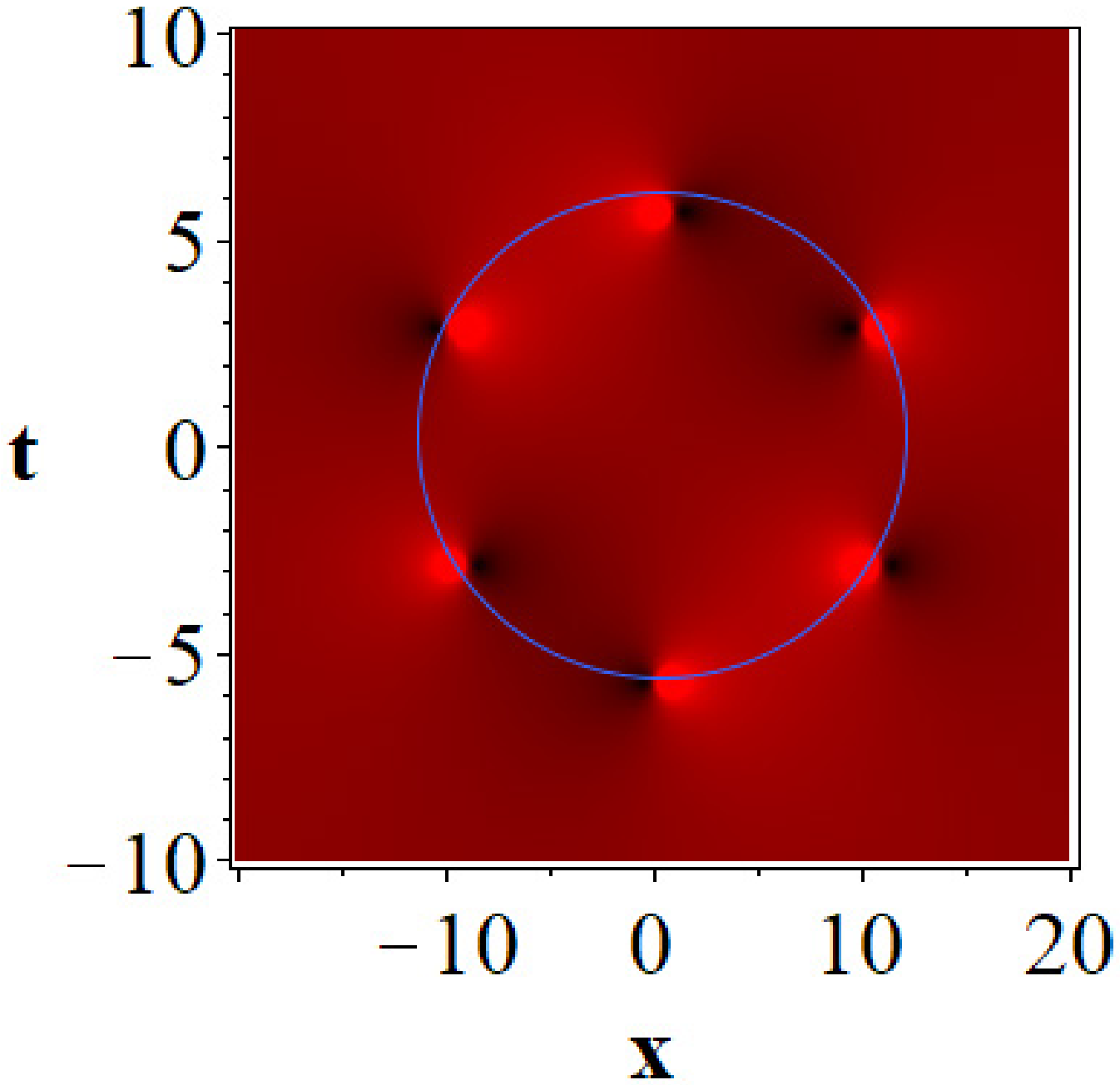}}}\\
$~~~~~~~~~~~~~~~~~~~~~~~~~~~~~~~~~~~~~~~~~~~~~~~~~~~~~~(\textbf{b})$\\

\noindent { \small \textbf{Figure 2.} Second-order rogue waves of a scalar nonlocal NLSE
with parameter values
$\rho=1, (\omega_{1,0},\omega_{2,0})=(1,0), (\omega_{1,1},\omega_{2,1})=(0,1000\textrm{i})$.\\}

$~~~~~~~~~~~~~~~~~~~$
{\rotatebox{0}{\includegraphics[width=8.2cm,height=4.8cm,angle=0]{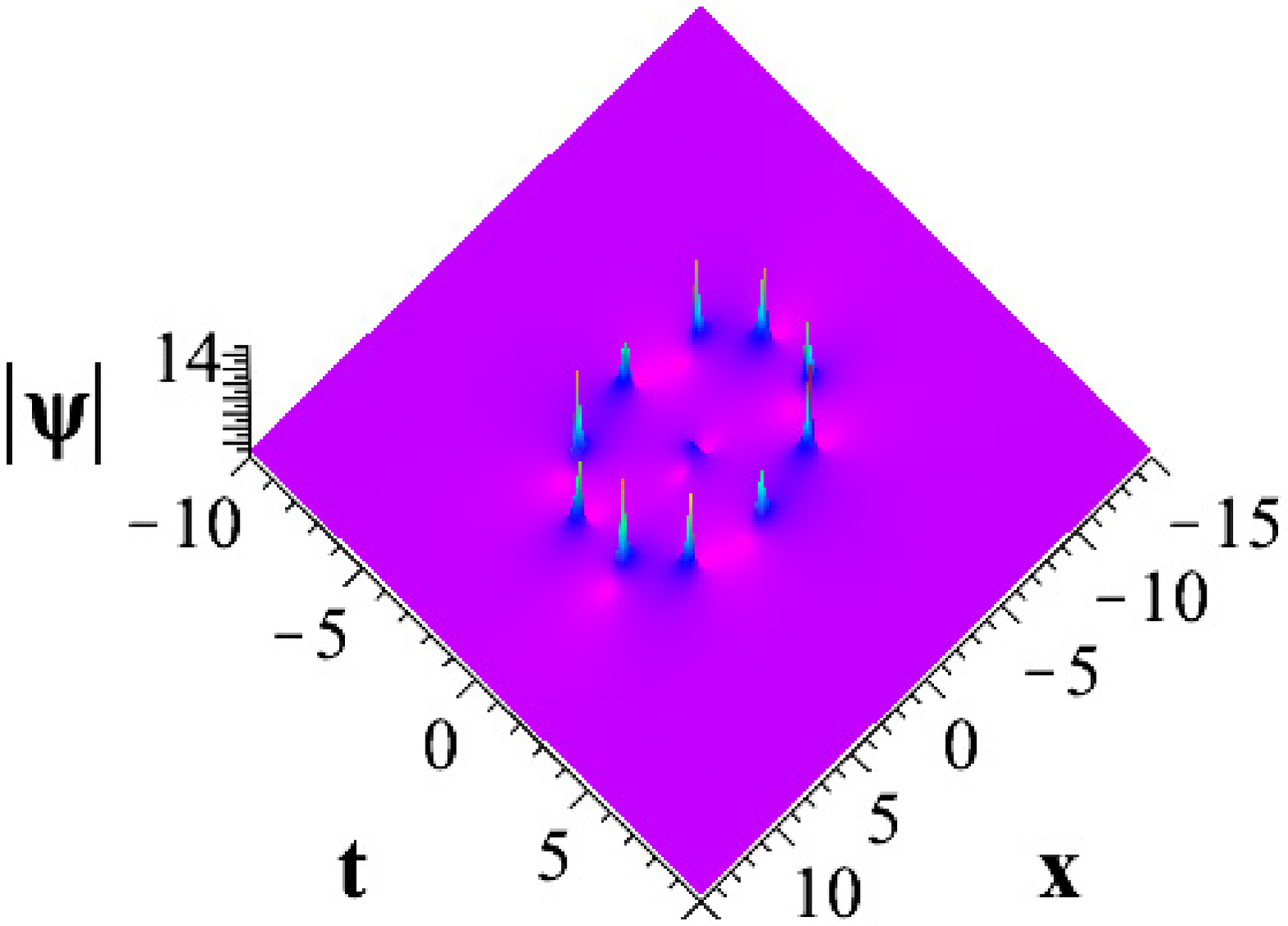}}}\\
$~~~~~~~~~~~~~~~~~~~~~~~~~~~~~~~~~~~~~~~~~~~~~~~~~~~~~~(\textbf{a})$\\

$~~~~~~~~~~~~~~~~~~~~~~~$
{\rotatebox{0}{\includegraphics[width=6.2cm,height=5.6cm,angle=0]{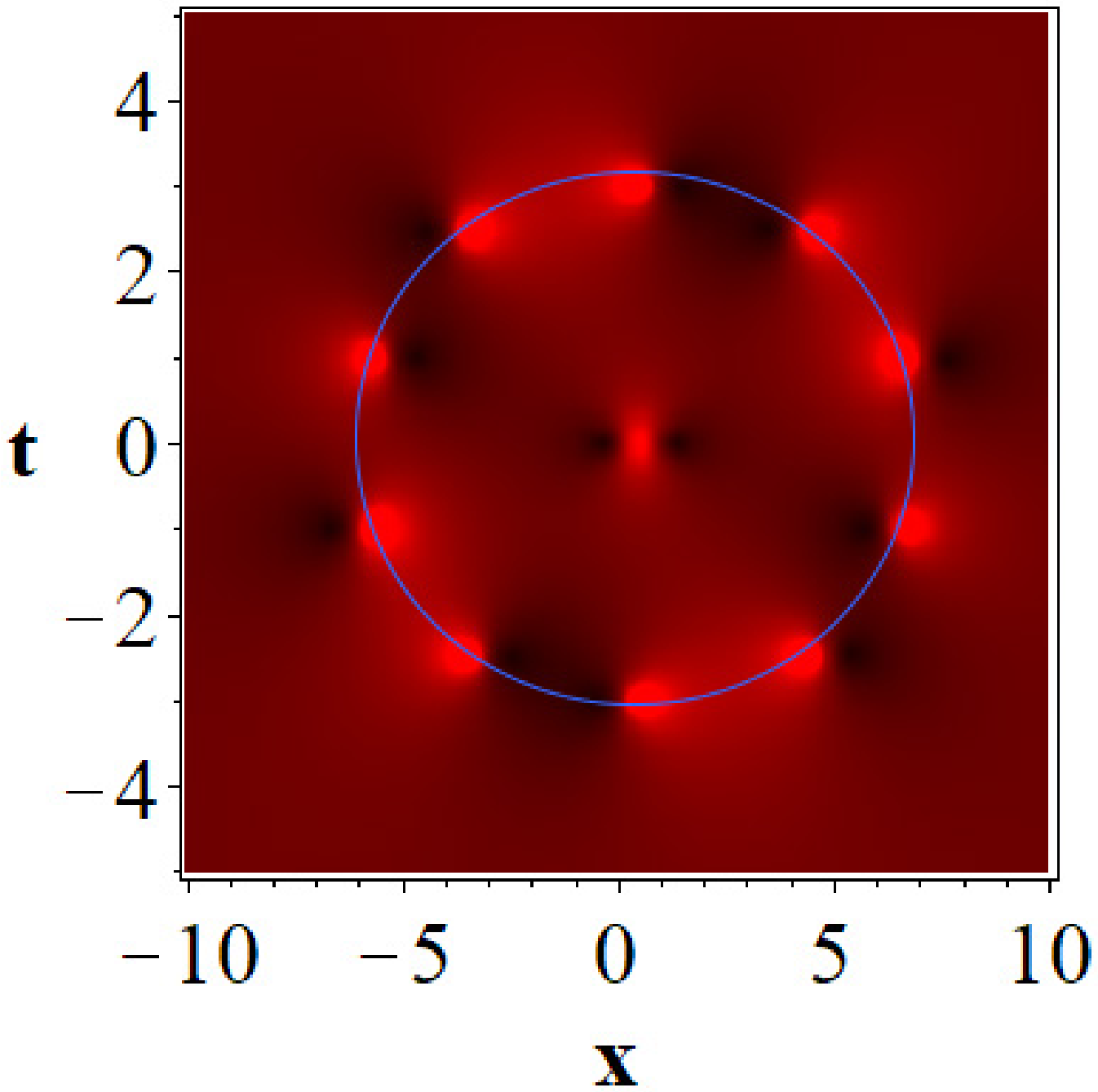}}}\\
$~~~~~~~~~~~~~~~~~~~~~~~~~~~~~~~~~~~~~~~~~~~~~~~~~~~~~~(\textbf{b})$\\

\noindent { \small \textbf{Figure 3.} Third-order rogue waves of a scalar nonlocal NLSE
with parameter values
$\rho=1, (\omega_{1,0},\omega_{2,0})=(1,0), (\omega_{1,1}=0,\omega_{2,1})=(0,0), (\omega_{1,2},\omega_{2,2})=(0,1000\textrm{i})$.\\}

$~~~~~~~~~~~~~~~~~~~$
{\rotatebox{0}{\includegraphics[width=8.2cm,height=4.8cm,angle=0]{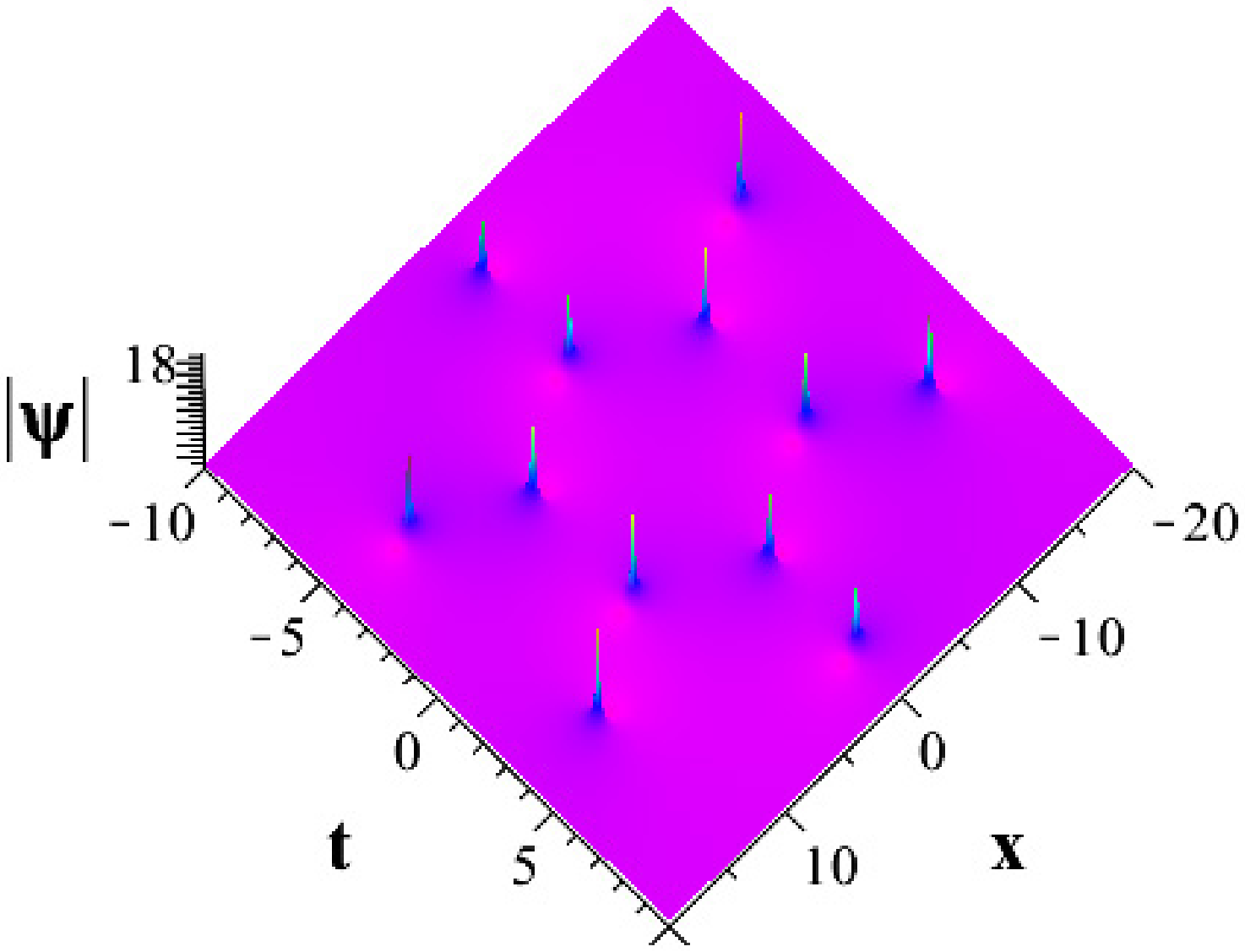}}}\\
$~~~~~~~~~~~~~~~~~~~~~~~~~~~~~~~~~~~~~~~~~~~~~~~~~~~~~~(\textbf{a})$\\

$~~~~~~~~~~~~~~~~~~~~~~~$
{\rotatebox{0}{\includegraphics[width=6.2cm,height=5.6cm,angle=0]{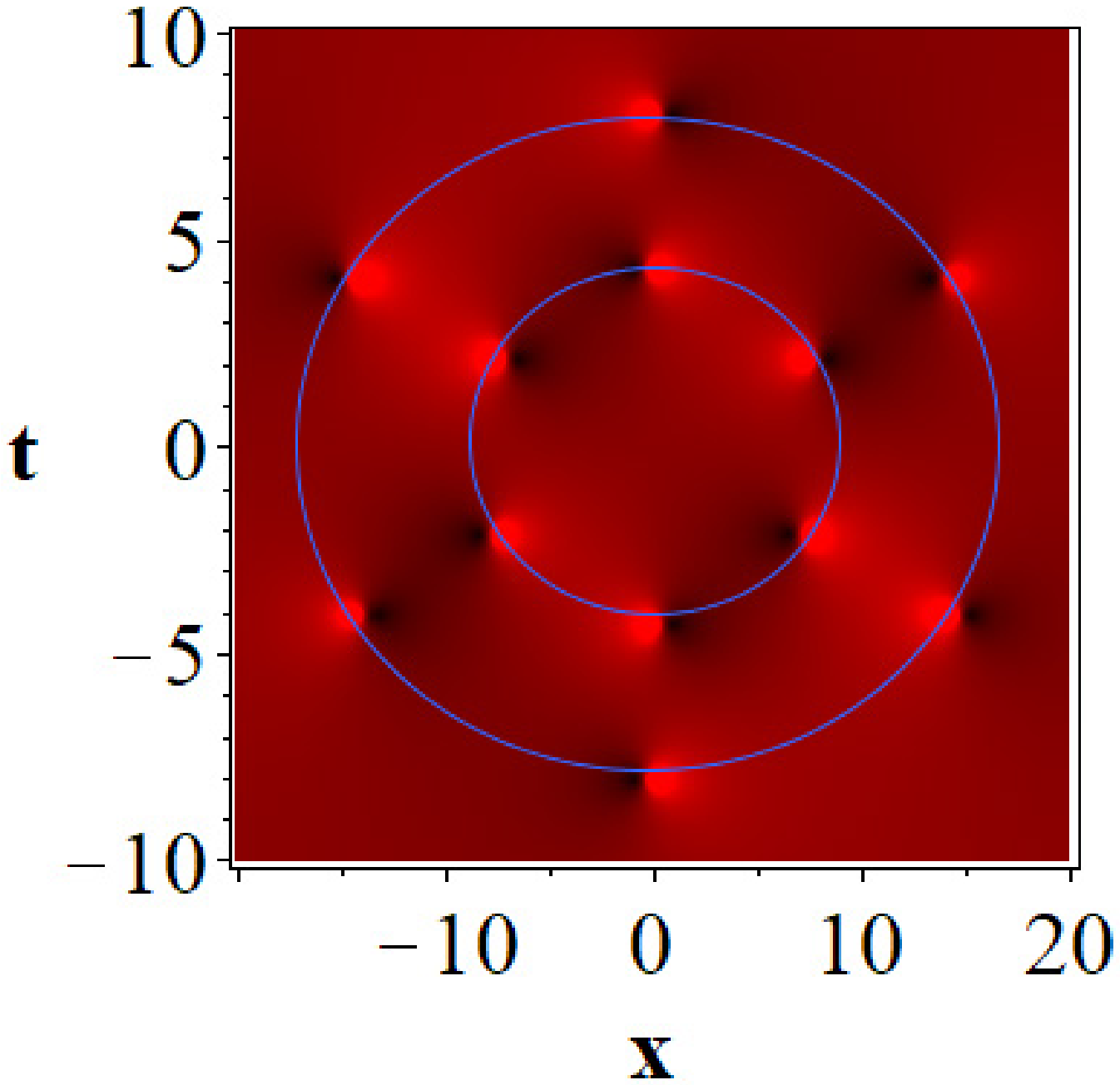}}}\\
$~~~~~~~~~~~~~~~~~~~~~~~~~~~~~~~~~~~~~~~~~~~~~~~~~~~~~~(\textbf{b})$\\

\noindent { \small \textbf{Figure 4.} Third-order rogue waves of a scalar nonlocal NLSE
with parameter values
$\rho=1, (\omega_{1,0},\omega_{2,0})=(1,1), (\omega_{1,1},\omega_{2,1})=(-1000\textrm{i},1000\textrm{i}), (\omega_{1,2},\omega_{2,2})=(0,1000\textrm{i})$.\\}

\section{A vector nonlocal nonlinear Schr\"{o}dinger equation}

As we all know, an increase in wave components can lead to some novel physical mechanism in nonlinear science.
Thus, to further enrich the nonlinear wave dynamics in vector models,
we next use the variable separation technique to derive the $N$th-order rogue wave
solutions of a vector nonlocal NLSE \eqref{RTNLSE2} based on DDT.

\subsection{Asymptotic expansion of Darboux-dressing transformation}

Eq.\eqref{RTNLSE2} admits the following Lax pair
\begin{equation}\label{1DDT-1}
\widetilde{\Psi}_{x}=\widetilde{\textbf{U}}\widetilde{\Psi},~~
\widetilde{\Psi}_{t}=\widetilde{\textbf{V}}\widetilde{\Psi},
\end{equation}
where
\begin{equation*}\label{1DDT-2}
\left\{ \begin{aligned}
&\widetilde{\textbf{U}}=\textrm{i}\lambda\widetilde{\sigma}_{3}+\widetilde{Q},\\
&\widetilde{\textbf{V}}=2\textrm{i}\lambda^2\widetilde{\sigma}_{3}+2\lambda \widetilde{Q}+\textrm{i}\widetilde{\sigma}_{3}\left(\widetilde{Q}^2-\widetilde{Q}_{x}\right),
                     \end{aligned} \right.
\end{equation*}
and
\begin{align*}\label{1DDT-3}
&\widetilde{Q}=\left[
    \begin{array}{ccc}
      0 & -\psi_{1}(x,-t) & -\psi_{2}(x,-t) \\
      \psi_{1}(x,t) & 0  & 0 \\
      \psi_{2}(x,t) & 0  & 0 \\
    \end{array}
  \right],\notag\\
  &\widetilde{\sigma}_{3}=\left[
                     \begin{array}{ccc}
                       1 & 0 & 0 \\
                       0 & -1 & 0 \\
                       0 & 0 & -1 \\
                     \end{array}
                   \right],
\end{align*}
with the spectral parameter $\lambda$.
By using the compatibility condition of system \eqref{1DDT-1}
\begin{equation*}\label{1CCSS1}
\widetilde{\textbf{U}}_{t}-\widetilde{\textbf{V}}_{x}+[\widetilde{\textbf{U}},\widetilde{\textbf{V}}]=0,
\end{equation*}
one can derive directly Eq.\eqref{RTNLSE2}.
The Lax pair \eqref{1DDT-1} admits the following symmetric condition
\begin{equation*}\label{1DDT-4}
\widetilde{\textbf{U}}(x,-t;-\lambda)=-\widetilde{\textbf{U}}^{\mbox{T}}(x,t;\lambda),~~\widetilde{\textbf{V}}(x,-t;-\lambda)=\widetilde{\textbf{V}}^{\mbox{T}}(x,t;\lambda).
\end{equation*}
We know that for $\lambda\in\mathbb{C}$, then $\widetilde{\Phi}_{1}=\widetilde{\Psi}_{1}^{\mbox{T}}(x,-t)$ solves the adjoint eigenvalue problem
\begin{equation*}\label{DDT-5}
\widetilde{\Phi}_{x}=-\widetilde{\Phi} \widetilde{\textbf{U}}(\widetilde{Q},\lambda),~~\widetilde{\Phi}_{t}=-\widetilde{\Phi} \widetilde{\textbf{V}}(\widetilde{Q},\lambda),
\end{equation*}
at $\lambda=-\lambda_{1}$, while $\widetilde{\Psi}_{1}$ is a solution of the linear matrix eigenvalue problem \eqref{1DDT-1} at $\lambda=\lambda_{1}$.

A suitable DDT for Eq.\eqref{RTNLSE2} is given by
\begin{equation}\label{1DDT-6}
\widetilde{\Psi}[1]=\widetilde{\textbf{D}}\widetilde{\Psi},
~~\widetilde{\textbf{D}}=\textbf{I}_{3\times3}-\frac{2\lambda_{1}}{\lambda+\lambda_{1}}\widetilde{\textbf{P}},~~
\widetilde{\textbf{P}}=\frac{\widetilde{\Delta}[0](x,t)\widetilde{\Delta}[0]^{\mbox{T}}(x,-t)}{\widetilde{\Delta}[0]^{\mbox{T}}(x,-t)\widetilde{\Delta}[0](x,t)},
\end{equation}
where $\textbf{I}_{3\times3}=\mbox{diag}(1,1,1)$, $\widetilde{\Delta}[0]=[\widetilde{\varphi}_{0},\widetilde{\varphi}_{1},\widetilde{\varphi}_{2}]^{\mbox{T}}$, and
$\Psi(x,t;\lambda_{1})$ is the fundamental solution for the Lax equations \eqref{1DDT-1} corresponding to $\lambda=\lambda_{1}$.
Next, we rewrite the DDT \eqref{1DDT-6} in the alternative form
\begin{equation}\label{1DDT-8}
\widetilde{\Psi}[1]=\widetilde{\textbf{T}}\widetilde{\Psi},~~\widetilde{\textbf{T}}=(\lambda+\lambda_{1})\textbf{I}_{3\times3}-2\lambda_{1}\widetilde{\textbf{P}}.
\end{equation}
Since $\widetilde{\textbf{T}}$ is also a DT of Eq.\eqref{RTNLSE2}, it follows that
\begin{equation}\label{1DDT-9}
\widetilde{\textbf{T}}_{x}+\widetilde{\textbf{T}}\widetilde{\textbf{U}}+\widetilde{\textbf{U}}_{1}\widetilde{\textbf{T}}=0,
~~\widetilde{\textbf{T}}_{t}+\widetilde{\textbf{T}}\widetilde{\textbf{U}}+\widetilde{\textbf{V}}_{1}\widetilde{\textbf{T}}=0.
\end{equation}
The matrices $\widetilde{\textbf{U}}_{1}$ and $\widetilde{\textbf{V}}_{1}$ are obtained by replacing $\widetilde{Q}$ with $\widetilde{Q}_{1}$ in $\widetilde{\textbf{U}}$ and $\widetilde{\textbf{V}}$, respectively.
It follows from \eqref{1DDT-8} and \eqref{1DDT-9}  that
\begin{equation*}\label{1DDT-10}
\widetilde{Q}_{1}=\widetilde{Q}_{0}+2\textrm{i}\lambda_{1}[\widetilde{\sigma}_{3},\widetilde{\textbf{P}}].
\end{equation*}
Here we note that
\begin{equation*}\label{1DDT-11}
\widetilde{\textbf{T}}|_{\lambda=\lambda_{1}}\widetilde{\Delta}[0]=0.
\end{equation*}
Similar to the results in the previous section, we present the following expansion theorem which can be used to derive new
solutions for the same spectral parameter.
\\

\noindent
\textbf{Theorem 3.1}
Let $\widetilde{\Psi}(\lambda)|_{\lambda=\lambda_{1}(1+\widetilde{\epsilon})}$  be a solution of the Lax system \eqref{1DDT-1} corresponding to the spectral parameter
$\lambda_{1}(1+\widetilde{\epsilon})$ and a seed solution $\widetilde{\psi}^{[0]}$.
Expanding $\widetilde{\Psi}(\lambda)$ at $\lambda_{1}$ by the Taylor expansion, we have
\begin{equation*}\label{DDT-12}
\widetilde{\Psi}(\lambda)|_{\lambda=\lambda_{1}(1+\epsilon)}=\widetilde{\Psi}_{0}\widetilde{\epsilon}+\Psi_{0}\widetilde{\epsilon}
+\widetilde{\Psi}_{0}\widetilde{\epsilon}^2+\cdots,
\end{equation*}
where
\begin{align*}\label{DDT-13}
&\widetilde{\Delta}[n]=\left(
            \begin{array}{c}
             \widetilde{ \varphi}_{0}^{[n]} \\
              \widetilde{\varphi}_{1}^{[n]} \\
              \widetilde{\varphi}_{2}^{[n]} \\
            \end{array}
          \right)=\lambda_{1}\widetilde{\Delta}[n-1]+\widetilde{\textbf{T}}[n]\widetilde{\Upsilon}[n-1],~~n\geq1,\notag\\
&\widetilde{\Delta}[0]=\left(
            \begin{array}{c}
              \widetilde{\varphi}_{0}^{[0]} \\
              \widetilde{\varphi}_{1}^{[0]} \\
              \widetilde{\varphi}_{2}^{[0]} \\
            \end{array}
          \right)=\widetilde{\Psi}_{0},\notag\\
&\widetilde{\Upsilon}[n-1]=\widetilde{\Delta}[n-1](\widetilde{\Psi}_{j}\rightarrow\widetilde{\Psi}_{j+1}),~~j=0,1,2,\cdots,
\end{align*}
with
\begin{equation*}\label{DDT-14}
\begin{aligned}
\widetilde{\textbf{T}}[n]=2\lambda_{1}(\textbf{I}_{3\times3}-\widetilde{\textbf{P}}[n]),\\
\widetilde{\textbf{P}}[n]=\frac{\widetilde{\Delta}[n-1](x,t)\widetilde{\Delta}^{\mbox{T}}[n-1](-x,t)}
{\widetilde{\Delta}^{\mbox{T}}[n-1](-x,t)\widetilde{\Delta}[n-1](x,t)},
                      \end{aligned}
\end{equation*}
are solutions of the Lax system \eqref{1DDT-1} corresponding to $\lambda_{1}$ and solution $\widetilde{\psi}^{[n]}$
\begin{align*}\label{DDT-15}
&\widetilde{\psi}^{[n]}=\widetilde{\psi}^{[n-1]}+4\textrm{i}\lambda_{1}\widetilde{\varphi}_{0}^{[n-1]}(x,-t)/\left[\widetilde{\varphi}_{0}^{[n-1]}(x,t)\widetilde{\varphi}_{0}^{[n-1]}(x,-t)\right.\notag\\
&\left.+\widetilde{\varphi}_{1}^{n-1}(x,t)\widetilde{\varphi}_{1}^{[n-1]}(x,-t)
+\widetilde{\varphi}_{2}^{n-1}(x,t)\widetilde{\varphi}_{2}^{[n-1]}(x,-t)\right]\notag\\
&\times\left[
                                                                                                                                                     \begin{array}{c}
                                                                                                                                                       \widetilde{\varphi}_{1}^{[n-1]}(x,t) \\
                                                                                                                                                       \widetilde{\varphi}_{2}^{[n-1]}(x,t) \\
                                                                                                                                                     \end{array}
                                                                                                                                                   \right]
.
\end{align*}


\subsection{The variable separation technique}

Following the same way, we choose the seed solutions of Eq.\eqref{RTNLSE2} as
\begin{equation}\label{1DDT-71}
\psi_{1}^{[0]}=\widetilde{\rho}\exp\left(2\textrm{i}\widetilde{\rho}^2 t\right),~~\psi_{2}^{[0]}=0,
\end{equation}
where $\widetilde{\rho}$ is free constant.
Then we find a family of the solutions of the Lax system \eqref{1DDT-71} corresponding to the spectral parameter $\lambda$ as follows
\begin{equation}\label{1DDT-81}
\widetilde{\Psi}=\left[
       \begin{array}{c}
         \widetilde{\varphi}_{0} \\
         \widetilde{\varphi}_{1} \\
         \widetilde{\varphi}_{2} \\
       \end{array}
     \right]=
\widetilde{\Lambda}\widetilde{\mathcal {R}}\widetilde{\mathcal {E}}\widetilde{\mathcal {Z}},
~~\widetilde{\mathcal {R}}=\exp(\textrm{i}\widetilde{\Theta} x),~~\widetilde{\mathcal {E}}=\exp(\textrm{i}\widetilde{\Omega} t),
\end{equation}
where
\begin{equation*}\label{1DDT-91}
\widetilde{\Lambda}=\left[
          \begin{array}{ccc}
            \exp\left(-\textrm{i}\widetilde{\rho}^2 t\right) & 0 & 0 \\
            0 & \exp\left(\textrm{i}\widetilde{\rho}^2 t\right) & 0 \\
            0 & 0 & \exp\left(\textrm{i}\widetilde{\rho}^2 t\right) \\
          \end{array}
        \right],
\end{equation*}
and $\widetilde{\mathcal {Z}}$ is a free complex vector.
Similar to the derivation of \eqref{DDT-12}, we obtain
\begin{equation*}\label{1DDT-12}
\widetilde{\Theta}=\left[
         \begin{array}{ccc}
           \lambda & \textrm{i}\widetilde{\rho} & 0 \\
           -\textrm{i}\widetilde{\rho} & -\lambda & 0 \\
           0 & 0 & -\lambda \\
         \end{array}
       \right],~~\widetilde{\Omega}=\widetilde{\Theta}^2+2\lambda\widetilde{\Theta}-\left(\lambda^2+\widetilde{\rho}^2\right).
\end{equation*}
Then the exponential matrices $\mathcal {R}$ and $\mathcal {E}$ in \eqref{1DDT-81} can be written as
\begin{equation}\label{1DDT-131}
\widetilde{\mathcal {R}}=\frac{1}{\widetilde{\tau}}\left[
                             \begin{array}{ccc}
                               \widetilde{\Theta}_{1} & -\widetilde{\Theta}_{2} & -\widetilde{\Theta}_{3} \\
                               \widetilde{\Theta}_{2} & \widetilde{\Theta}_{4} & \widetilde{\Theta}_{5} \\
                               \widetilde{\Theta}_{3} & \widetilde{\Theta}_{5} & \widetilde{\Theta}_{6} \\
                             \end{array}
                           \right],~~
\widetilde{\mathcal {E}}=\frac{1}{\widetilde{\xi}}\left[
                             \begin{array}{cccc}
                               \widetilde{\Omega}_{1} & -\widetilde{\Omega}_{2} & -\widetilde{\Omega}_{3} \\
                               \widetilde{\Omega}_{2} & \widetilde{\Omega}_{4} & \widetilde{\Omega}_{5} \\
                               \widetilde{\Omega}_{3} & \widetilde{\Omega}_{5} & \widetilde{\Omega}_{6} \\
                             \end{array}
                           \right],
\end{equation}
where
\begin{align*}\label{1DDT-141}
&\widetilde{\Theta}_{1}=\widetilde{\tau}\cos(\widetilde{\tau} x)+\textrm{i}\lambda\sin(\widetilde{\tau} x),~~
\widetilde{\Theta}_{2}=\widetilde{\rho}\sin(\widetilde{\tau} x),~~\widetilde{\Theta}_{3}=\widetilde{\Theta}_{5}=0, \notag\\
&\widetilde{\Theta}_{4}=\widetilde{\tau}\cos(\widetilde{\tau} x)-\textrm{i}\lambda\sin(\widetilde{\tau} x),~~
\widetilde{\Theta}_{6}=\widetilde{\tau} \exp(-\textrm{i}\lambda x),\notag\\
&\widetilde{\Omega}_{1}=\widetilde{\xi}\cos(\widetilde{\xi} t)+2\textrm{i}\lambda^2\sin(\widetilde{\xi} t),~~
\widetilde{\Omega}_{2}=2\lambda\widetilde{\rho} \sin(\widetilde{\xi} t),~~\widetilde{\Omega}_{3}=\widetilde{\Omega}_{5}=0,\notag\\
&\widetilde{\Omega}_{4}=\widetilde{\xi}\cos(\widetilde{\xi} t)-2\textrm{i}\lambda^2\sin(\widetilde{\xi} t),~~\widetilde{\Theta}_{6}=\xi\exp\left(-\textrm{i}\widetilde{\rho}^2t-2\textrm{i}\lambda^2t\right),\notag\\
&\widetilde{\xi}=2\lambda\widetilde{\tau},~~\widetilde{\tau}=\sqrt{\lambda^2+\widetilde{\rho}^2}.
\end{align*}

\subsection{Construction of $N$th-order rogue wave solutions}

In this subsection, we seek $N$th-order rogue waves of Eq.\eqref{RTNLSE2} using the Theorem 3.1 presented in the previous subsection.
Taking $\lambda=\textrm{i}\widetilde{\rho}(1+\widetilde{\epsilon})$ in \eqref{1DDT-131}.
Then using Taylor series expansions for the trigonometric and exponential functions,
the matrix $\widetilde{\mathcal {R}}$ has the following expansion at $\widetilde{\epsilon}=0$
\begin{equation*}\label{1DDT-15}
\widetilde{\mathcal {R}}\big|_{\lambda=\textrm{i}\widetilde{\rho}(1+\widetilde{\epsilon})}=\sum_{n=1}^{\infty}\widetilde{\mathcal {R}}_{n}\widetilde{\epsilon}^n,
\end{equation*}
where
\begin{equation*}\label{1DDT-16}
\mathcal {R}_{n}=\left[
                            \begin{array}{ccc}
                              \widetilde{\alpha}_{n}-\widetilde{\beta}_{n}-\widetilde{\beta}_{n-1} & -\widetilde{\beta}_{n} & 0 \\
                              \widetilde{\beta}_{n} & \widetilde{\alpha}_{n}+\widetilde{\beta}_{n}+\widetilde{\beta}_{n-1} & 0 \\
                               0 & 0 & \exp(\widetilde{\rho} x)\widetilde{\textbf{A}}_{n}\\
                            \end{array}
                          \right],
\end{equation*}
with
\begin{equation*}\label{DDT-17}
\left\{ \begin{aligned}
&\widetilde{\alpha}_{n}=\sum_{l=0}^{\left[\frac{n}{2}\right]}\textbf{C}_{n-l}^{l}2^{n-2l}\widetilde{\textbf{A}}_{2(n-l)},\\
&\widetilde{\beta}_{n}=\sum_{l=0}^{\left[\frac{n}{2}\right]}\textbf{C}_{n-l}^{l}2^{n-2l}\widetilde{\textbf{A}}_{2(n-l)+1},\\
&\textbf{C}_{n}^{m}=\frac{n!}{m!(n-m)!},
~~\widetilde{\textbf{A}}_{m}=\frac{\widetilde{\rho}^mx^{m}}{m!},~~n\geq m,~~m,n\in\mathbb{N}^{+}.
                     \end{aligned} \right.
\end{equation*}
Following the same way, the matrix $\widetilde{\mathcal {G}}$ has the following expansion at $\widetilde{\epsilon}=0$
\begin{equation*}\label{1DDT-18}
\widetilde{\mathcal {E}}\big|_{\lambda=\textrm{i}\rho(1+\widetilde{\epsilon})}=
\sum_{n=0}^{\infty}\widetilde{\mathcal {E}}_{n}\widetilde{\epsilon}^{n},
\end{equation*}
where
\begin{equation*}\label{1DDT-19}
\widetilde{\mathcal {E}}_{n}=\left[
                            \begin{array}{ccc}
                              \widetilde{\gamma}_{n}-\textrm{i}\widetilde{\theta}_{n}-\textrm{i}\widetilde{\theta}_{n-1} & -\widetilde{\gamma}_{n} & 0  \\
                              \widetilde{\gamma}_{n} & \widetilde{\gamma}_{n}+\textrm{i}\widetilde{\theta}_{n}+\textrm{i}\widetilde{\theta}_{n-1}  & 0 \\
                              0 & 0  & \exp\left(\textrm{i}\widetilde{\rho}^2 t\right)\rho_{n}
                            \end{array}
                          \right],
\end{equation*}
with
\begin{equation*}\label{1DDT-20}
\left\{ \begin{aligned}
&\widetilde{\gamma}_{n}=\sum_{l=0}^{\left[\frac{3n}{4}\right]}\sum_{m=0}^{l}(-1)^{n-l}\textbf{C}_{n-l}^{m}
\textbf{C}_{2(n-l)}^{l-m}2^{n-l-m}\widetilde{\textbf{B}}_{2(n-l)},\\
&\widetilde{\theta}_{n}=\sum_{l=0}^{\left[\frac{3n+1}{4}\right]}\sum_{m=0}^{l}(-1)^{n-l}\textbf{C}_{n-l}^{m}
\textbf{C}_{2(n-l)+1}^{l-m}2^{n-l-m}\widetilde{\textbf{B}}_{2(n-l)+1},\\
&\rho_{n}=\sum_{l=0}^{[n/2]}\textbf{C}_{n-l}^{l}\textrm{i}^{n-l}2^{n-2l}\widetilde{\textbf{B}}_{n},\\
&\widetilde{\textbf{B}}_{m}=\frac{\widetilde{\rho}^{2m}2^{m}t^{m}}{m!},~~l\in\mathbb{N}^{+}.
                     \end{aligned} \right.
\end{equation*}
Let us next assume $\widetilde{\omega}_{k}$ to be an arbitrary polynomial function of $\widetilde{\epsilon}$ given by
\begin{equation}\label{1DDT-21}
\widetilde{\mathcal {Z}}_{0}(\widetilde{\epsilon})=\sum_{k=0}^{n}\widetilde{\omega}_{k}\widetilde{\epsilon}^k,~~
\widetilde{\omega}_{k}=\left[
                      \begin{array}{c}
                        \widetilde{\omega}_{1,k} \\
                        \widetilde{\omega}_{2,k}\\
                        \widetilde{\omega}_{3,k}\\
                      \end{array}
                    \right],
\end{equation}
thus solution \eqref{1DDT-8} has an expansion
\begin{align*}\label{1DDT-22}
\widetilde{\Psi}\big|_{\lambda=\textrm{i}\widetilde{\rho}(1+\widetilde{\epsilon})}=\sum_{n=0}^{\infty}\widetilde{\Psi}_{n}\widetilde{\epsilon}^{n},~~
\widetilde{\Psi}_{n}=\widetilde{\Lambda}\sum_{k=0}^{n}
\sum_{j=0}^{n}\widetilde{\mathcal {F}}_{k}\widetilde{\mathcal {G}}_{j}\widetilde{\omega}_{n-k-j}.
\end{align*}
Here we rewrite $\widetilde{\omega}_{k}$ in \eqref{1DDT-21} in a new form
\begin{equation*}\label{new-1}
 \sum_{k=0}^{\infty}\widetilde{\omega}_{k}\widetilde{\epsilon}^{k}=\exp\left(\textrm{i}
 \widetilde{\Theta}|_{\lambda=\textrm{i}\widetilde{\rho}(1+\widetilde{\epsilon})}x_{0}
 +\textrm{i}\Omega|_{\lambda=\textrm{i}\widetilde{\rho}(1+\widetilde{\epsilon})}t_{0}\right)\widetilde{l},
\end{equation*}
where
\begin{align*}\label{new-2}
x_{0}=r_{0}+r_{1}\widetilde{\epsilon}+r_{2}\widetilde{\epsilon}^{2}+\ldots,~~t_{0}=s_{0}+s_{1}\widetilde{\epsilon}+s_{2}\widetilde{\epsilon}^{2}+\ldots,
\end{align*}
and $\widetilde{l}=(l_{1},l_{2},l_{3})^{\mbox{T}}$.
Taking $\lambda_{1}=\textrm{i}\widetilde{\rho}$ in Theorem 3.1, we then obtain the $N$th-order rogue wave solutions of Eq.\eqref{RTNLSE2}.

%

In what follows, we will discuss the dynamic behaviors of the rogue wave solutions in the framework of Eq.\eqref{RTNLSE2} by graphic representations.

$~~~~~~~~~~~~~~~~~~~$
{\rotatebox{0}{\includegraphics[width=7.8cm,height=4.8cm,angle=0]{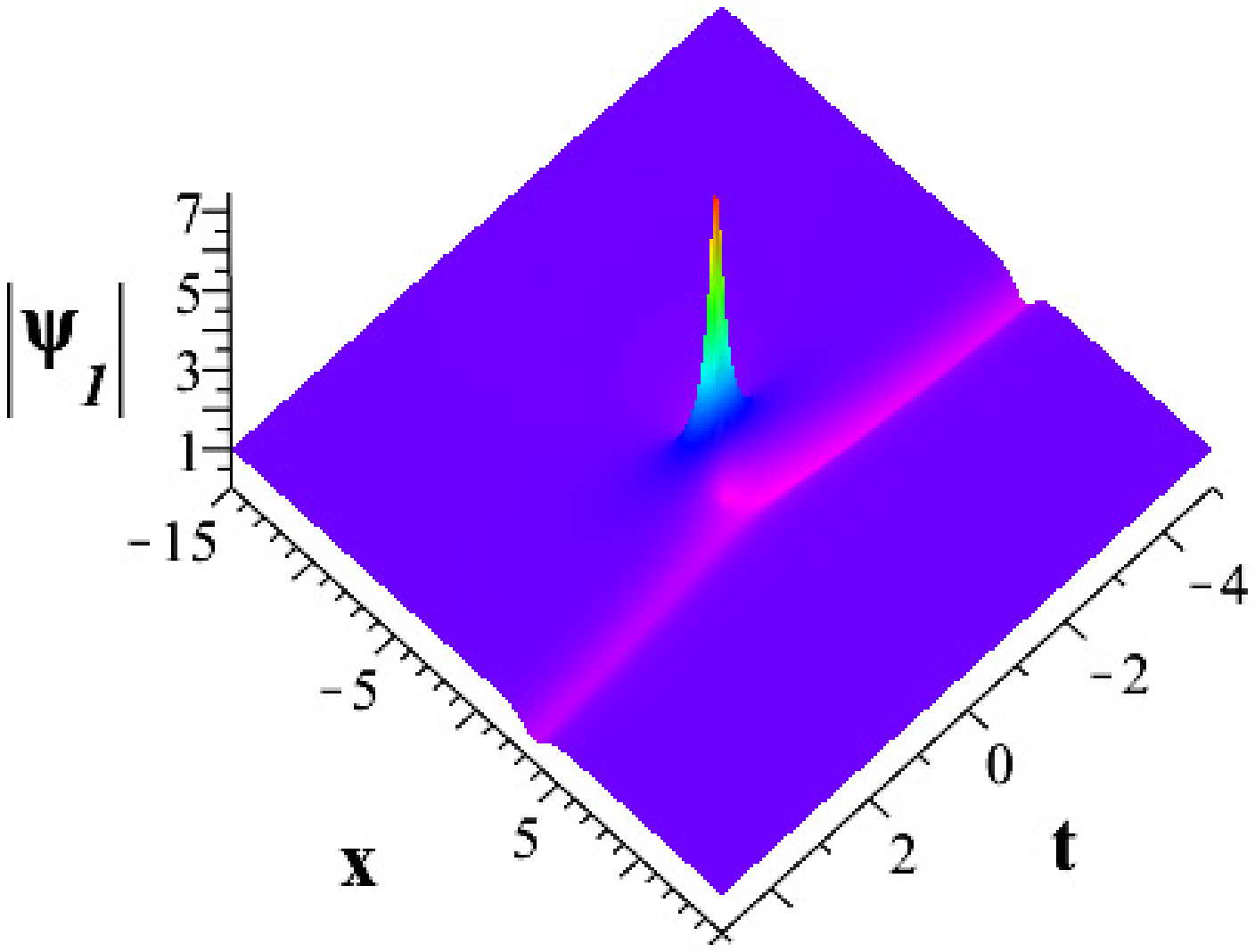}}}\\
$~~~~~~~~~~~~~~~~~~~~~~~~~~~~~~~~~~~~~~~~~~~~~~~~~~~~~~(\textbf{a})$\\

$~~~~~~~~~~~~~~~~~~~~~~~$
{\rotatebox{0}{\includegraphics[width=6.2cm,height=5.6cm,angle=0]{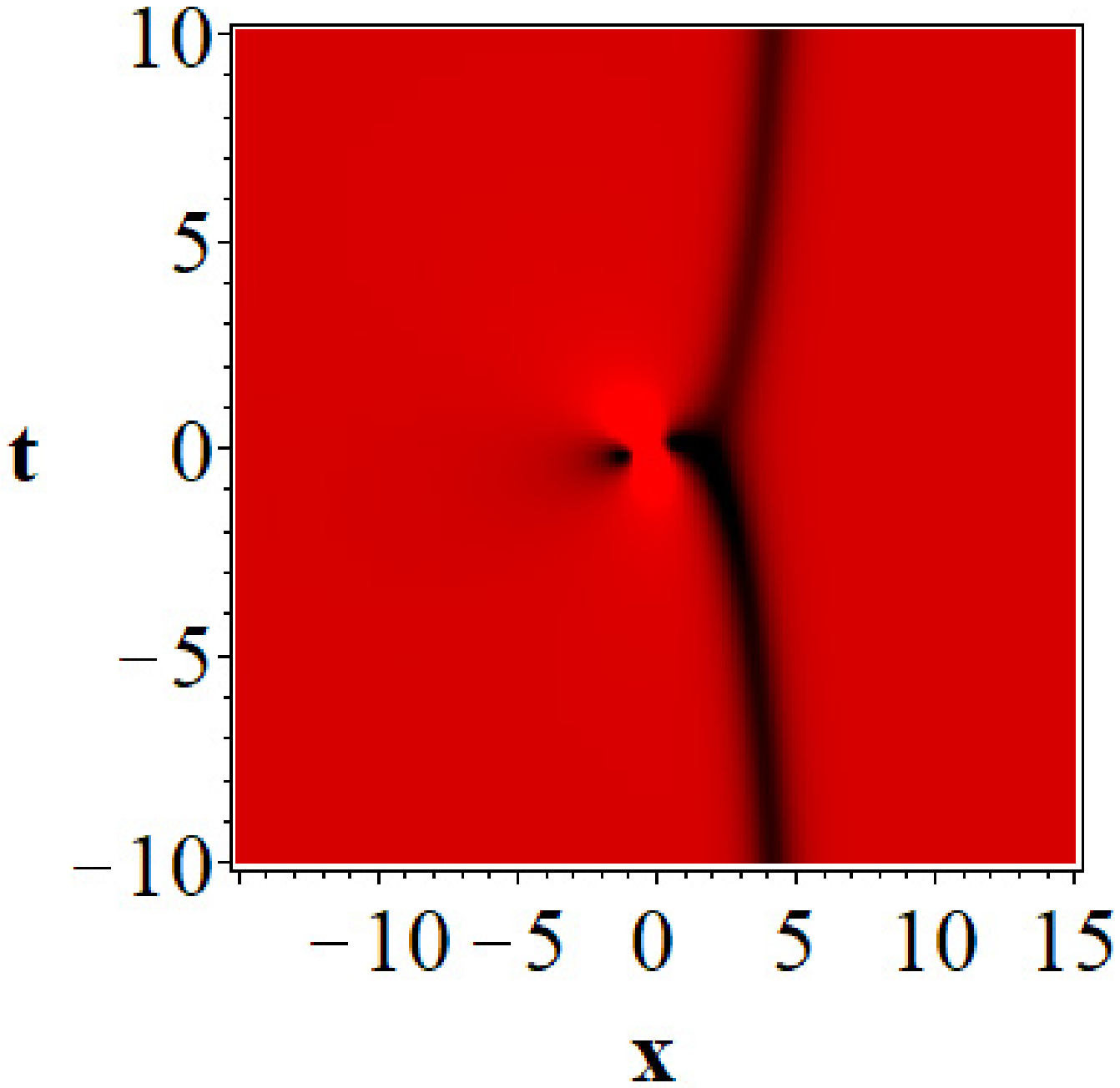}}}\\
$~~~~~~~~~~~~~~~~~~~~~~~~~~~~~~~~~~~~~~~~~~~~~~~~~~~~~~(\textbf{b})$\\

$~~~~~~~~~~~~~~~~~~~$
{\rotatebox{0}{\includegraphics[width=7.8cm,height=4.8cm,angle=0]{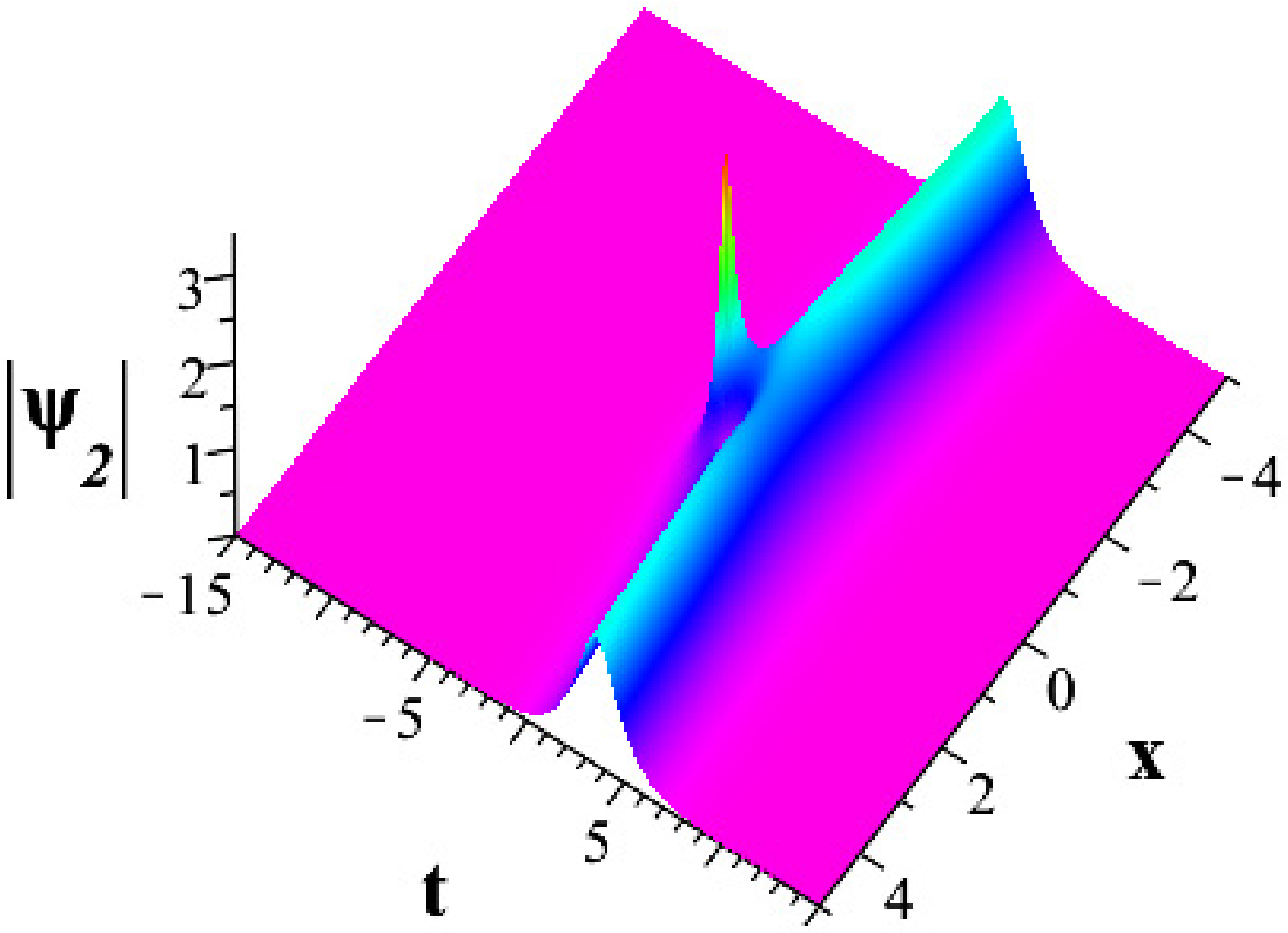}}}\\
$~~~~~~~~~~~~~~~~~~~~~~~~~~~~~~~~~~~~~~~~~~~~~~~~~~~~~~~~(\textbf{c})$\\

$~~~~~~~~~~~~~~~~~~~~~~~$
{\rotatebox{0}{\includegraphics[width=6.2cm,height=5.6cm,angle=0]{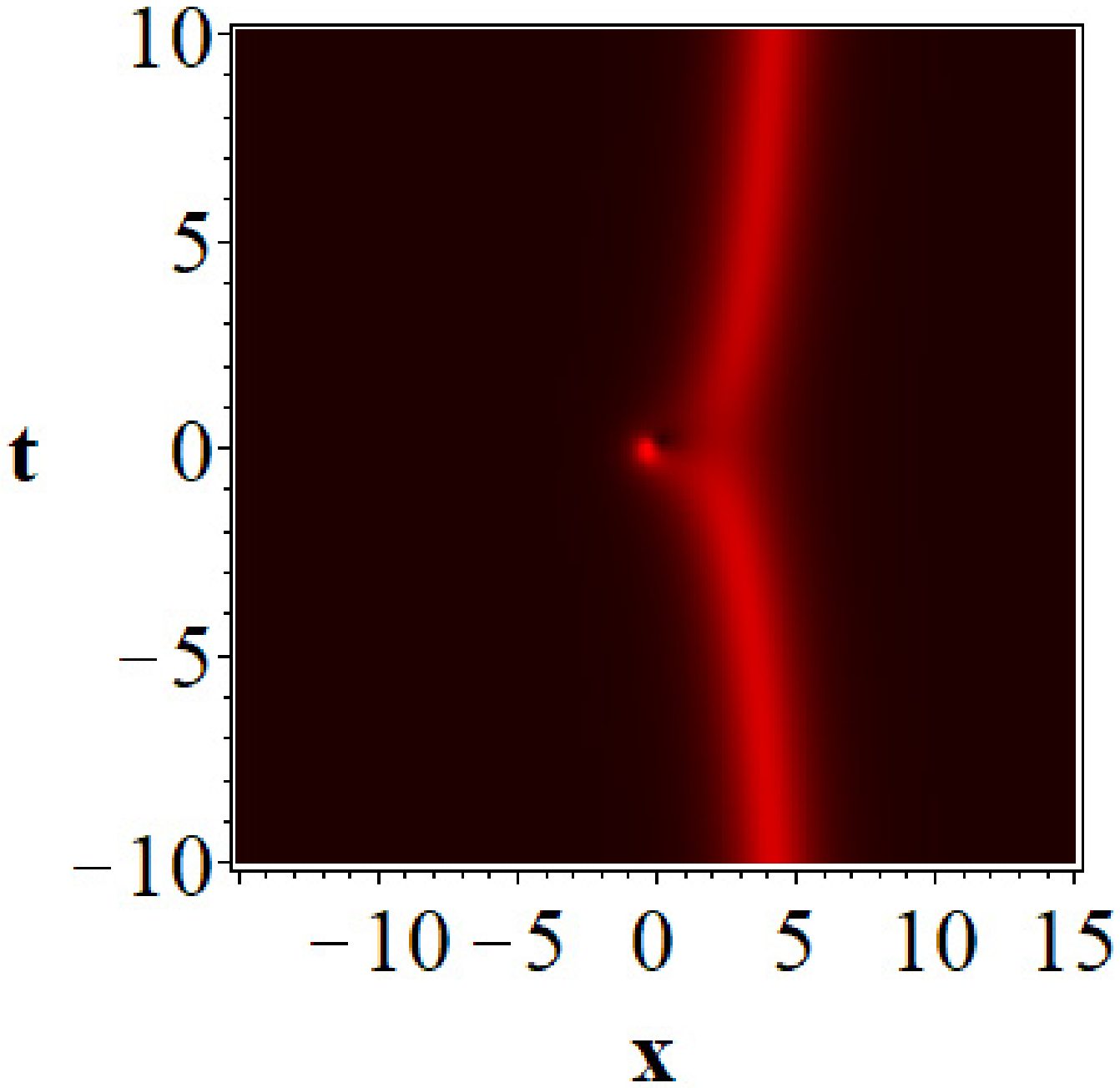}}}\\
$~~~~~~~~~~~~~~~~~~~~~~~~~~~~~~~~~~~~~~~~~~~~~~~~~~~~~~~~(\textbf{d})$\\

\noindent { \small \textbf{Figure 5.} First-order rogue wave of a vector nonlocal NLSE
with parameter values
$\rho=1, (\widetilde{\omega}_{1,0},\widetilde{\omega}_{2,0},\widetilde{\omega}_{3,0})=(1,2\textrm{i},\textrm{i})$.\\}

$~~~~~~~~~~~~~~~~~~~$
{\rotatebox{0}{\includegraphics[width=7.8cm,height=4.8cm,angle=0]{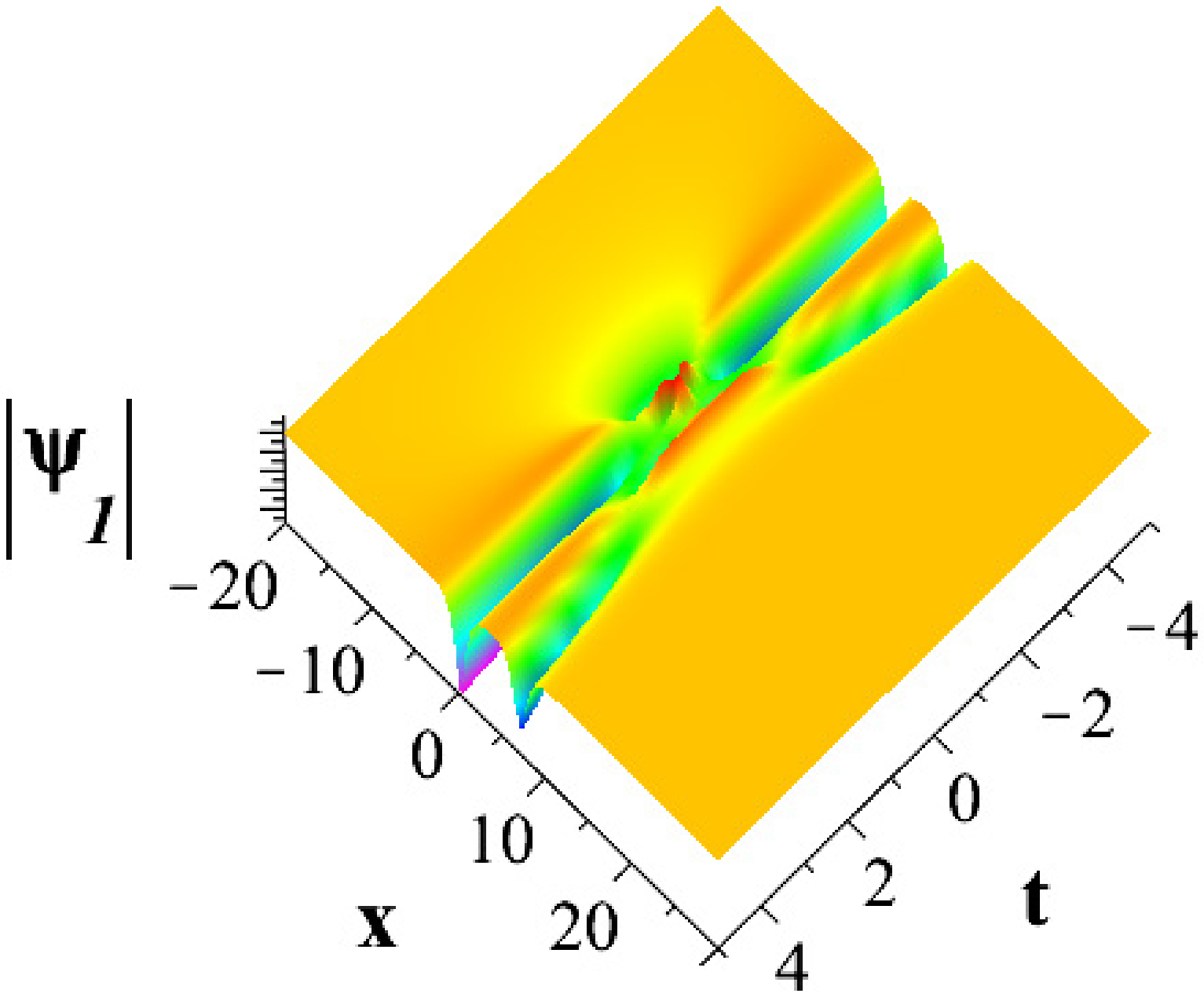}}}\\
$~~~~~~~~~~~~~~~~~~~~~~~~~~~~~~~~~~~~~~~~~~~~~~~~~~~~~~(\textbf{a})$\\

$~~~~~~~~~~~~~~~~~~~~~~~$
{\rotatebox{0}{\includegraphics[width=6.2cm,height=5.6cm,angle=0]{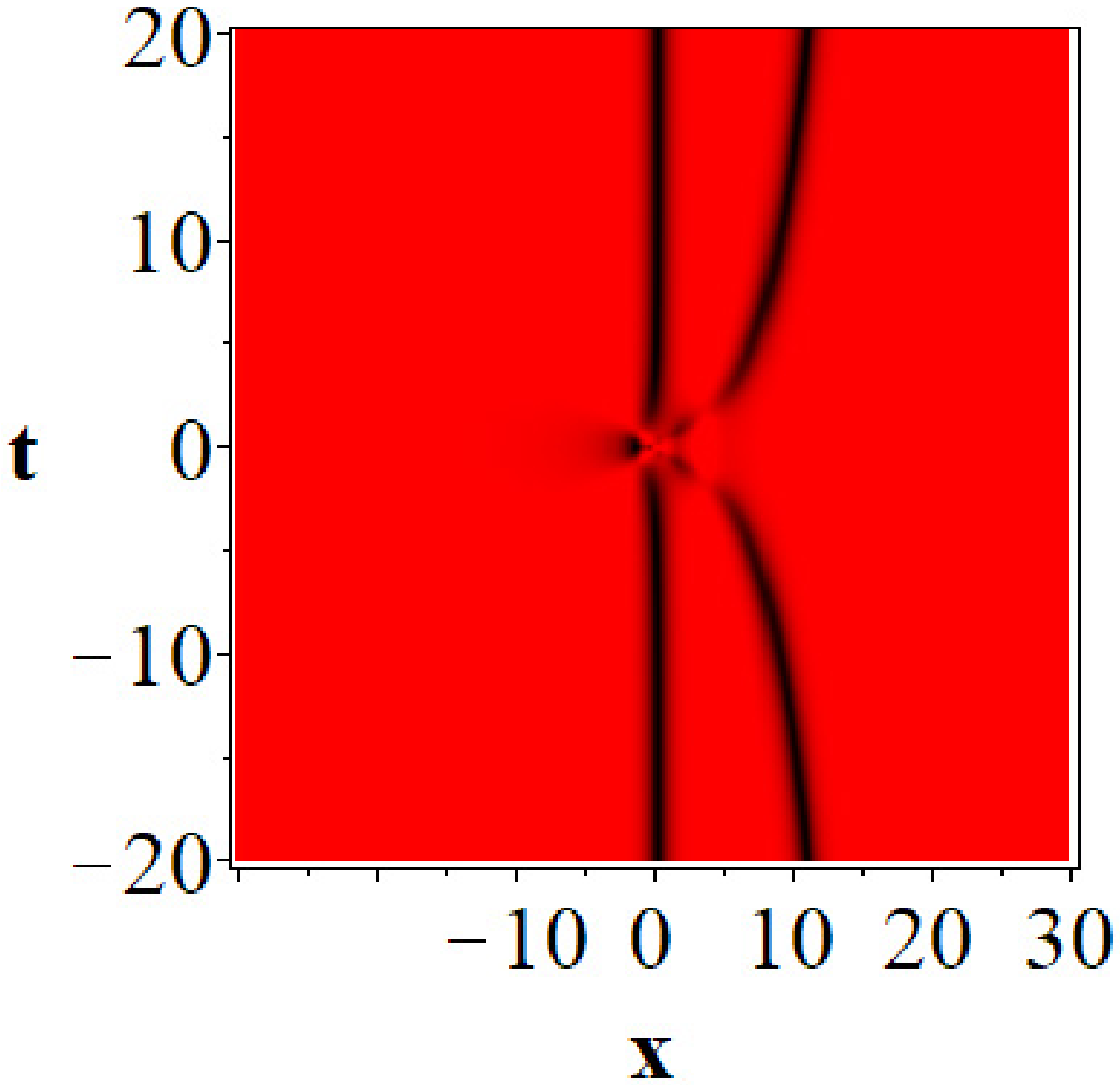}}}\\
$~~~~~~~~~~~~~~~~~~~~~~~~~~~~~~~~~~~~~~~~~~~~~~~~~~~~~~(\textbf{b})$\\

$~~~~~~~~~~~~~~~~~~~$
{\rotatebox{0}{\includegraphics[width=7.8cm,height=4.8cm,angle=0]{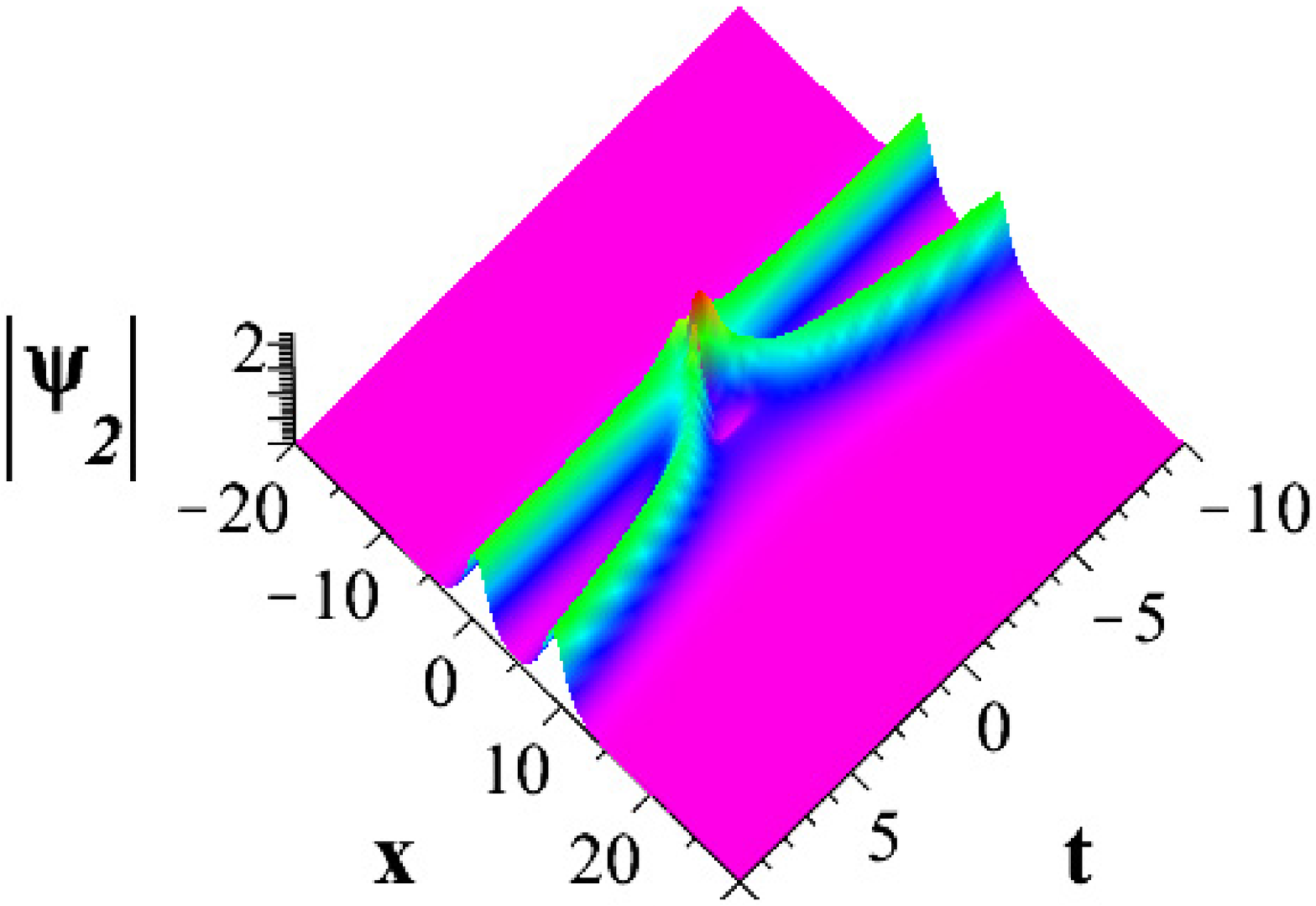}}}\\
$~~~~~~~~~~~~~~~~~~~~~~~~~~~~~~~~~~~~~~~~~~~~~~~~~~~~~~~~(\textbf{c})$\\

$~~~~~~~~~~~~~~~~~~~~~~~$
{\rotatebox{0}{\includegraphics[width=6.2cm,height=5.6cm,angle=0]{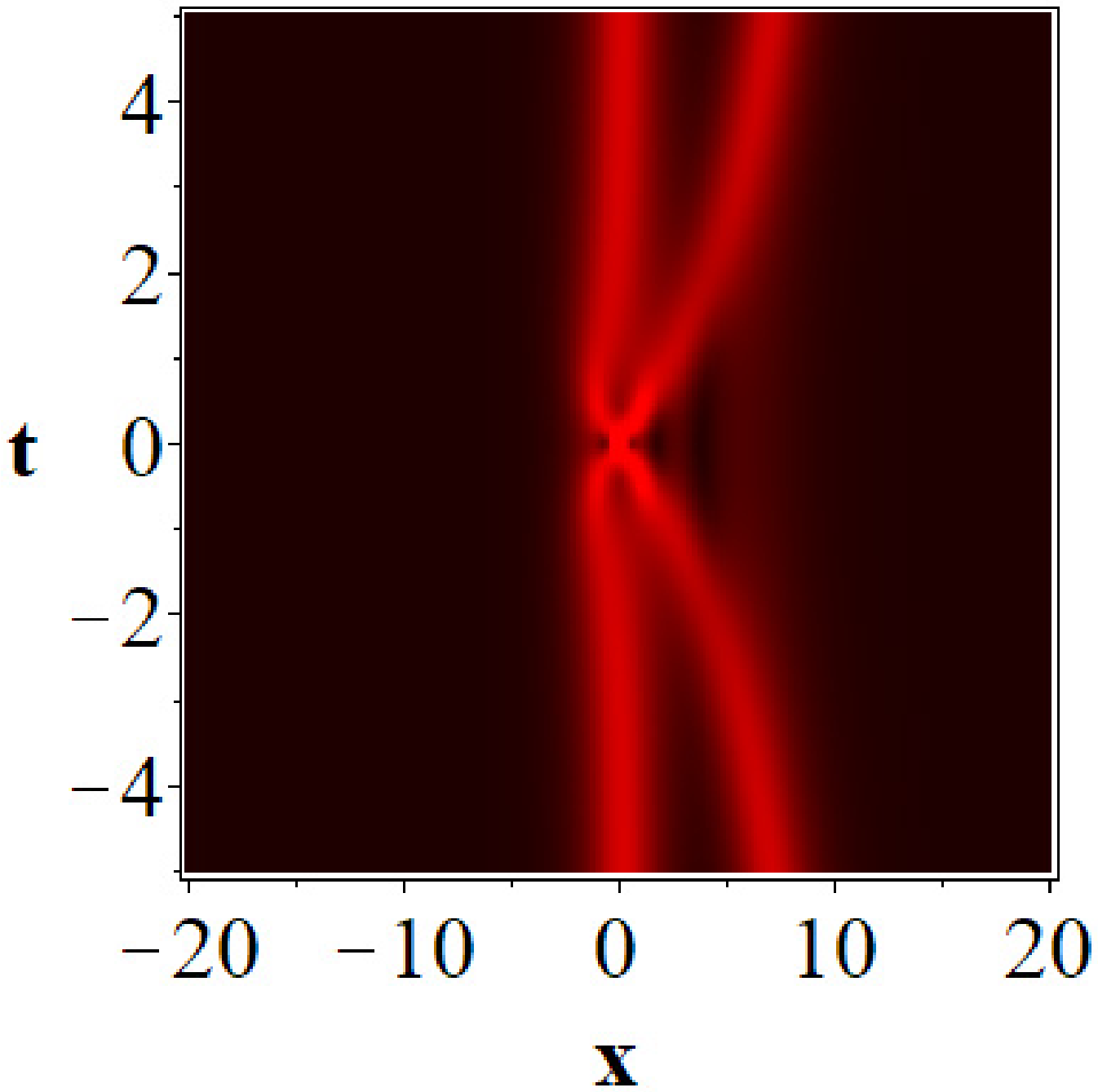}}}\\
$~~~~~~~~~~~~~~~~~~~~~~~~~~~~~~~~~~~~~~~~~~~~~~~~~~~~~~~~(\textbf{d})$\\

\noindent { \small \textbf{Figure 6.} Second-order rogue waves of a vector nonlocal NLSE
with parameter values
$\widetilde{\rho}=1, (l_{1},l_{2},l_{3})=(5\times10^{7},5\times10^{7},1)$, $r_{j}=0, s_{j}=0$ for all $j$.\\}

(I)
Taking $N=1$, we have the first-order rogue wave solution of Eq.\eqref{RTNLSE2}.
In this case, one-peak-two-valleys rogue wave with a bright-dark soliton can be obtained.
As observed in Figure 5, we easily observe that $|\psi_{1}|$ and $|\psi_{2}|$ have different structures.
In the $|\psi_{1}|$ component, the one-peak-two-valleys rogue wave with a dark soliton is displayed in Figure 5(a),
while in the $|\psi_{2}|$ component, the one-peak-two-valleys rogue wave with a bright soliton appears, as shown in Figure 5(b).

(II)
Taking $N=2$, we have the second-order rogue wave solutions of Eq.\eqref{RTNLSE2}.
In this case, we see that two solitons and a second-order rogue wave coexist.
As observed in Figure 6, under the condition that the vector $r_{j}=0$ and $s_{j}=0$,
two bright (or dark) solitons together with a fundamental second-order rouge wave is displayed, and
the center of the rogue wave locate at the origin.
In Figure 7, if the other values keep unchanged and increase the values $s_{1}$, $s_{2}$,
in the $|\psi_{1}|$ component, the second-order rogue wave locate at the origin split into six singular peaks,
and this case gives rise to the two solitons that are far away from the rogue wave,
while the second-order rogue wave in the $|\psi_{2}|$ component is difficult to observe.

$~~~~~~~~~~~~~~~~~~~$
{\rotatebox{0}{\includegraphics[width=7.8cm,height=4.8cm,angle=0]{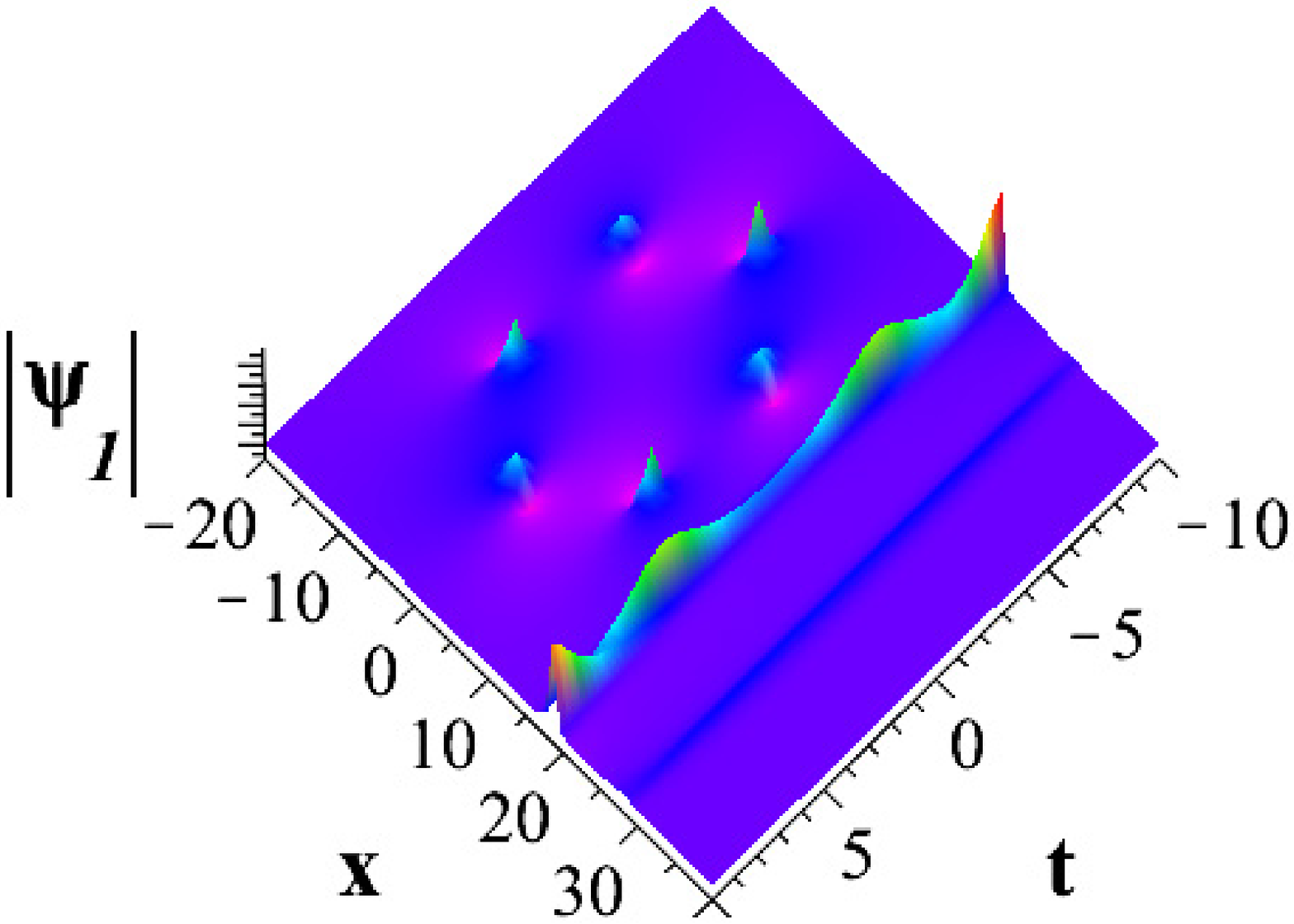}}}\\
$~~~~~~~~~~~~~~~~~~~~~~~~~~~~~~~~~~~~~~~~~~~~~~~~~~~~~~(\textbf{a})$\\

$~~~~~~~~~~~~~~~~~~~~~~~$
{\rotatebox{0}{\includegraphics[width=6.2cm,height=5.6cm,angle=0]{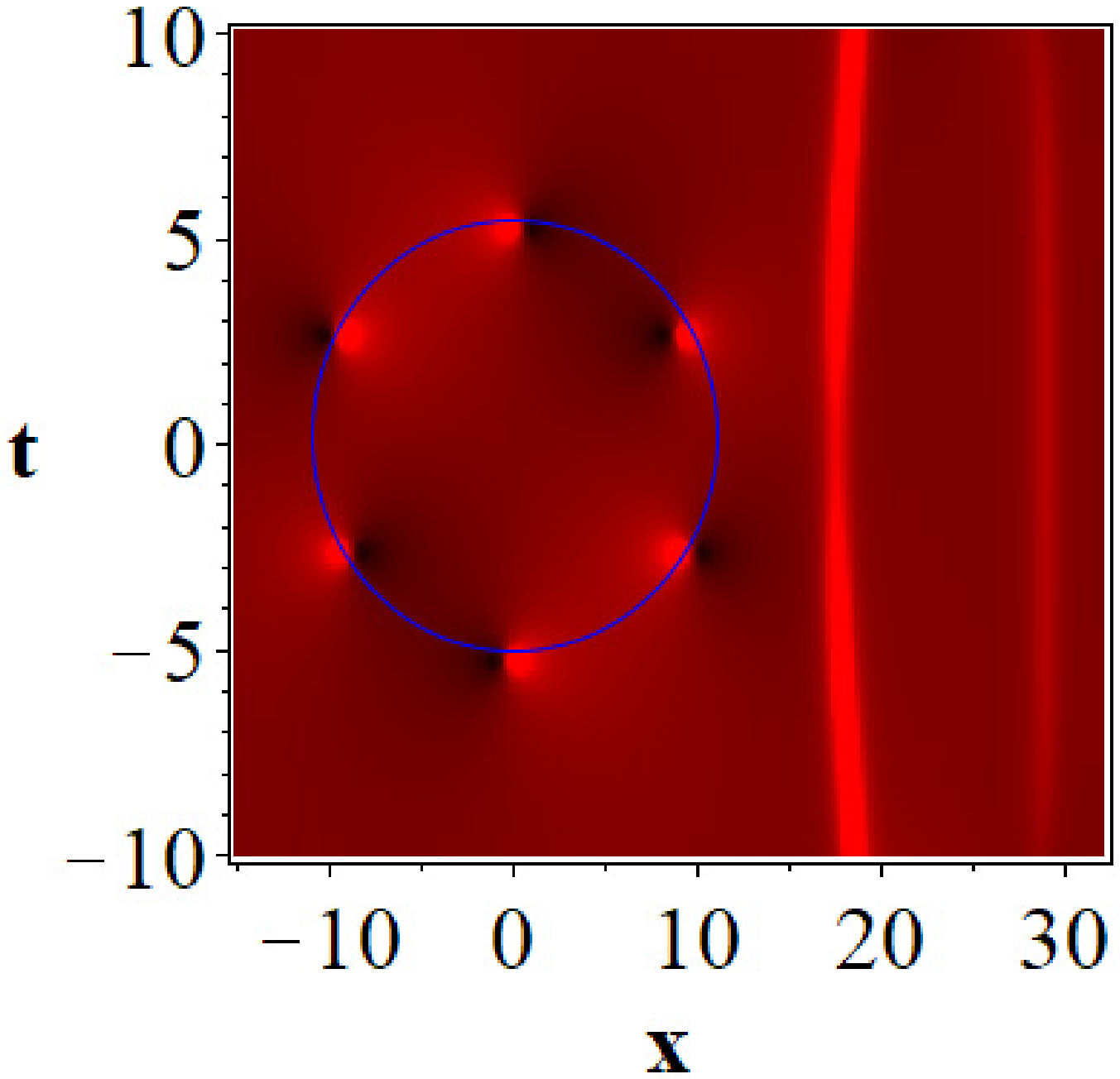}}}\\
$~~~~~~~~~~~~~~~~~~~~~~~~~~~~~~~~~~~~~~~~~~~~~~~~~~~~~~(\textbf{b})$\\

$~~~~~~~~~~~~~~~~~~~$
{\rotatebox{0}{\includegraphics[width=7.8cm,height=4.8cm,angle=0]{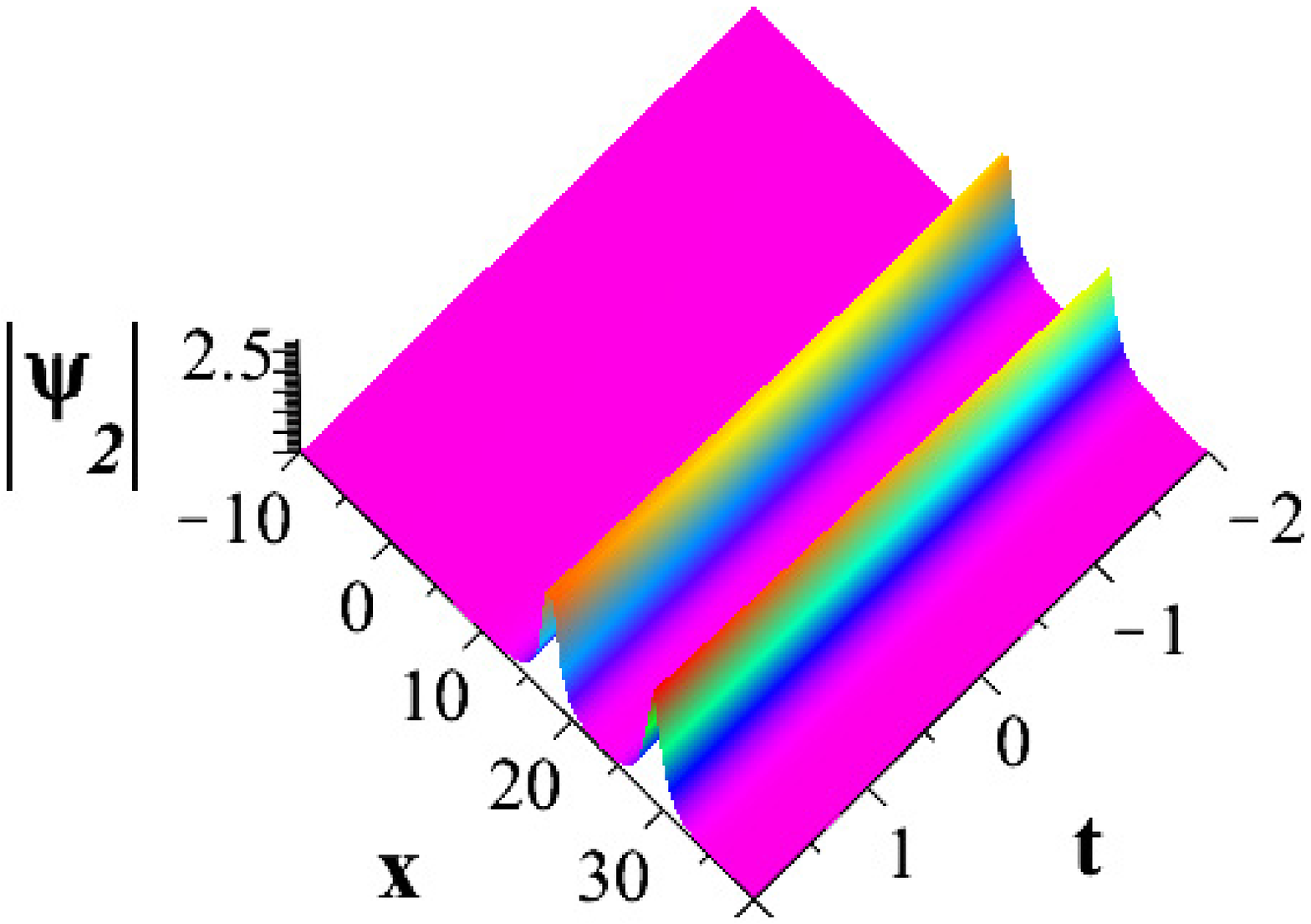}}}\\
$~~~~~~~~~~~~~~~~~~~~~~~~~~~~~~~~~~~~~~~~~~~~~~~~~~~~~~~~(\textbf{c})$\\

$~~~~~~~~~~~~~~~~~~~~~~~$
{\rotatebox{0}{\includegraphics[width=6.2cm,height=5.6cm,angle=0]{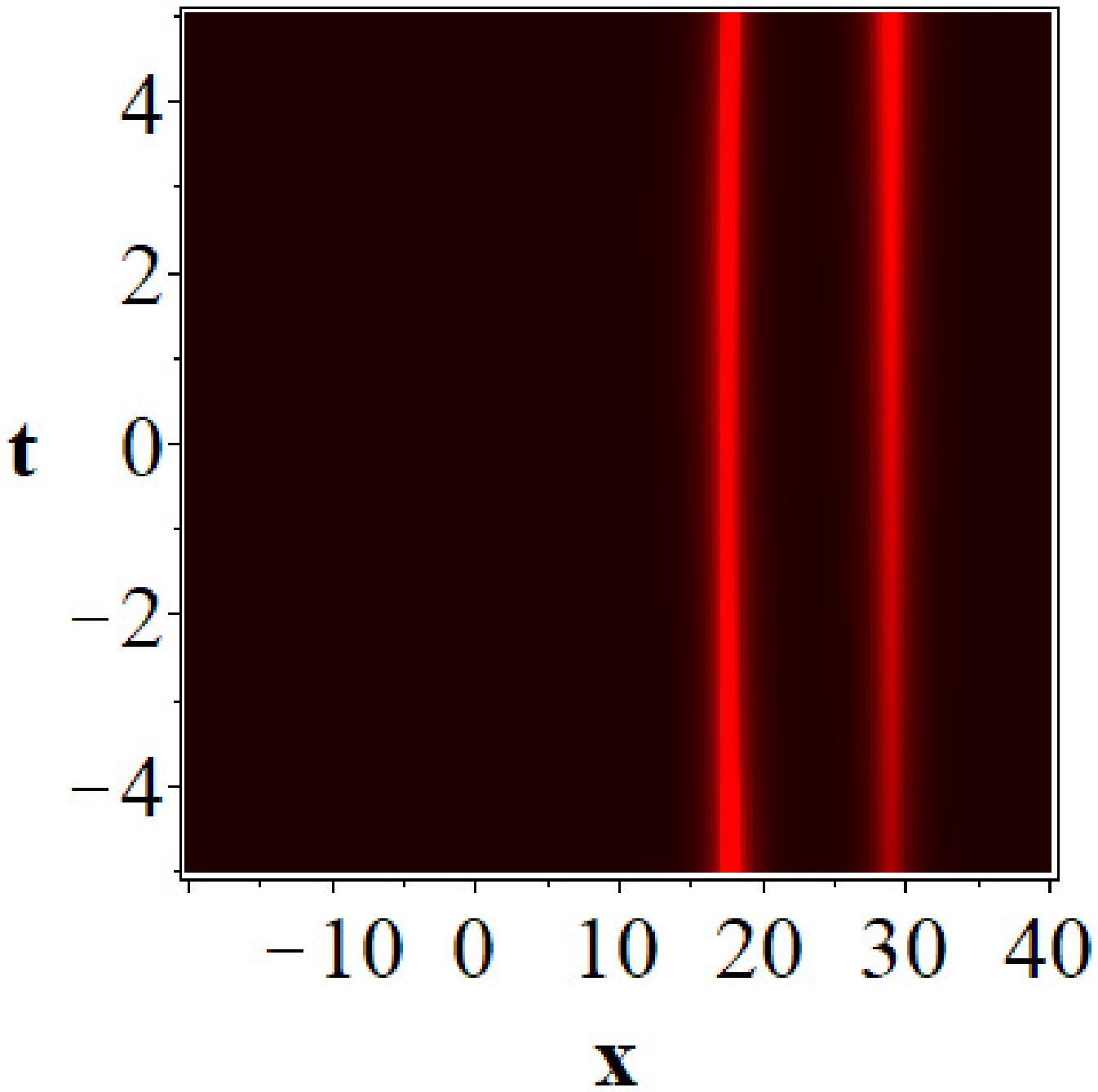}}}\\
$~~~~~~~~~~~~~~~~~~~~~~~~~~~~~~~~~~~~~~~~~~~~~~~~~~~~~~~~(\textbf{d})$\\

\noindent { \small \textbf{Figure 7.} Second-order rogue waves of a vector nonlocal NLSE
with parameter values
$\widetilde{\rho}=1, (l_{1},l_{2},l_{3})=(5\times10^{7},5\times10^{7},1)$, $r_{j}=0, s_{1}=0, s_{1}=400, s_{2}=300$ for all $j$.\\}

\section{Conclusions}

In this work, we have derived $N$th-order rogue wave
solutions of Eq.\eqref{RTNLSE} and Eq.\eqref{RTNLSE2} through a DDT by a separation of variable approach.
Moreover, the interesting and complicated dynamic patterns of these rogue waves have been discussed by varying the available parameters.
More interesting are the collapsing solutions, which show more complex patterns
which have not been observed in the corresponding local NLSEs.
In particular, comparing with the scalar nonlocal NLSE,
we find that the structure of rogue waves in vector nonlocal NLSE can exhibit rogue waves on a multisoliton background.
Moreover, under certain conditions, we can also observe ring structures of $N$th-order rogue waves on an $N$ bright-dark soliton background.
Although our explicit solutions exhibited here are lowest order, a parallel way can be used
to work out the $N$th-order rogue waves. Finally, it is worthy to emphasize that the technique presented in this
work may be available to construct rogue waves of matrix versions of the reverse-time nonlocal NLSE, even its hierarchy.
Additionally, these
results demonstrate that more abundant and novel rogue waves may exist in the nonlocal nonlinear equations than in the corresponding local ones.

\section*{Acknowledgements}
\hspace{0.3cm}
This work is supported by the NSFC under Grant
Nos. 12201622 and 11975306.






\end{document}